\documentclass[pre, aps, preprint, superscriptaddress]{revtex4-1}
\pdfoutput=1
\usepackage{amsmath,natbib,amsfonts}
\usepackage{amssymb,color, float}
\usepackage{graphicx}
\usepackage[centerlast]{caption}
\usepackage{subfig}
\usepackage[normalem]{ulem}
\usepackage{tikz}
\usepackage[colorlinks=true, citecolor=blue, urlcolor = blue, linkcolor= red,bookmarks=true]{hyperref}
\newcommand{\beqa}{\begin{eqnarray}}
\newcommand{\eeqa}{\end{eqnarray}}

\usepackage{multirow}
\usepackage{listings}
\usepackage[most]{tcolorbox}
\usepackage{xcolor}
\tcbset{
  colback=red!10!white,
  colframe=red!50!white,
  coltitle=black,
  fonttitle=\bfseries,
  title={Title Box},
  rounded corners,
  boxrule=0.5mm,
  arc=4mm,
  outer arc=2mm,
  top=1mm,bottom=1mm,left=8mm,right=1mm
}
\begin{document}
\title{Optimizing power and efficiency of a single spin heat engine}
\author{Rita Majumdar}
\thanks{These authors contributed equally.}
\affiliation{Department of Physics, Indian Institute of Technology Delhi, Hauz Khas 110016, New Delhi, INDIA}
\affiliation{ICTP -- The Abdus Salam International Centre for Theoretical Physics,  34151 Trieste, Italy}
\author{Monojit Chatterjee}
\thanks{These authors contributed equally.}
\affiliation{Department of Physics, Indian Institute of Technology Delhi, Hauz Khas 110016, New Delhi, INDIA}
\author{Rahul Marathe}
\email{maratherahul@physics.iitd.ac.in}
\affiliation{Department of Physics, Indian Institute of Technology Delhi, Hauz Khas 110016, New Delhi, INDIA}
\date{\today}
\begin{abstract}
We study the behavior of a single spin in the presence of a time-varying magnetic field utilizing Glauber dynamics. We engineer the system to function as an engine by changing the magnetic field according to specific protocols. Subsequently, we analyze the engine's performance using various protocols and stochastic thermodynamics to compute average values of crucial quantities for quantifying engine performance.  In the longtime limit of the engine cycle, we derive exact analytical expressions for work, heat, and efficiency in terms of a generalized protocol. We then analyze the model in terms of optimization of efficiency and power.
Additionally, we use different protocols and employ a gradient descent algorithm to best fit those to obtain optimal efficiency and then optimal power for a finite cycle time. All the protocols converge to the piece-wise constant protocol during efficiency optimization. We then explore a more general approach using the variational principle to determine the optimal protocols for optimizing power and efficiency. During the optimization process for both power and efficiency, the net entropy production decreases, which enhances the engine's performance. This approach demonstrates the superior optimization of efficiency and power in this system compared to the gradient descent algorithm.
\end{abstract}

\maketitle  

\section{Introduction}
Stochastic thermodynamics, an emerging field at the intersection of statistical physics and thermodynamics, has recently witnessed remarkable advancements that promise to reshape our understanding of the physical world. While traditional thermodynamics deals with the behavior of macroscopic systems governed by deterministic laws, stochastic thermodynamics works in the microscopic realm, where thermal fluctuations and probabilistic behavior play a pivotal role. Over the last decade, significant progress has been achieved towards understanding and describing fluctuation's role in such microscopic systems. The fluctuation theorem, the Jarzynski equality, Crook's theorem, and the formulation of stochastic thermodynamics provides a novel framework to tackle the role of fluctuation in entropy production and dissipative work for a system far from equilibrium~\cite{sekimoto98,Seifert12}. For stochastic processes, it is very crucial to understand energy conversion in  microscopic heat engines at the microscopic scales. Microscopic heat engines operate on scales where thermal fluctuations dominate and can extract useful work from these fluctuations \cite{Martinez17}. Significant research has been conducted on stochastic heat engines that utilize Brownian particles within active and passive media, as discussed in ~\cite{ Marathe18, Marathe19, Marathe22,Rana14,Ruben23}. Numerous studies have been focused on investigating spin systems from both classical and quantum mechanical perspectives as potential heat engines \cite{Marathe05,Marathe07,Marathe17,Chen02,Myers22}. 

In addition to theoretical research, substantial experimental efforts have been conducted within the realm of Brownian heat engines~\cite{Bechinger12,Martinez16,Krishnamurty16, Albay21, Cheng22} and spin systems~\cite{Robnagel16,Assis19,Klimovsky18,Barontini19,Levy20}. Single-atom heat engines have been accomplished by confinement of an ion within a linear Paul trap where the ion is thermally driven by alternatingly coupling it to hot and cold reservoirs~\cite{Robnagel16}. Additionally, research has explored quantum heat engines utilizing a spin-$1/2$ system and nuclear magnetic resonance techniques, operating within a regime where thermal and quantum fluctuations play a significant role~\cite{Peterson19}. An experimental study of a single ion confined in a linear Paul trap with tapered geometry coupled to engineered laser reservoirs to run the engine at maximum power has also been done \cite{Abah12}. Quantum engines considering quantum particles as a working substance can also deliver maximum power with maximum efficiency simultaneously \cite{Bera22}.
In numerous studies of heat engines, researchers try to identify the most efficient protocols through analytical and numerical methods \cite{Xiao14,Abiuso20}. In the context of Brownian systems, well-established techniques exist for determining the optimal protocol by optimizing work, heat, efficiency, and power, as documented in references~\cite{Seifert07,Udo08, Puglisi21, Holubec18}. Notably, numerical approaches have emerged where the Gradient Descent algorithm is applied to achieve maximum efficiency protocols for overdamped Brownian systems~\cite{Ye22,Fu2023,Dechant16}. For single-level quantum dots, analytical methods in the weak dissipation limit have been derived for optimal protocols by evaluating efficiency at maximum power~\cite{Massimo10,MassimoPRL10}. Recently, numerical techniques have been introduced to optimize both efficiency and power.  The implementation of these optimized protocols leads to enhanced heat engine performance. One such method involves utilizing the Pareto front to minimize a single-objective linear function through stochastic gradient descent~\cite{Solon18}. Also, optimization of a quantum dot heat engine using reinforcement learning is documented \cite{Erdman23}. The heat engines with classical spins and their optimization have not yet been explored extensively.

In this paper we consider a single spin under the influence of a magnetic field and interacting with two thermal baths at different temperatures. The strength of the magnetic field is varied periodically over time according to a fixed protocol that mimics a Stirling engine-like cycle. Here, a single classical spin acts like a working substance. We calculate the average values of different thermodynamic quantities in various cycle duration regimes. We analyze the spin dynamics and the evolution of the magnetization using Glauber dynamics \cite{Glauber,Marathe05}. We provide the essential mathematical formulas for computing various averaged thermodynamic quantities. We first consider a particular type of protocol where the maximum and minimum values of the time-dependent magnetic field are adjusted. Our objective is to explore the possible enhancement in efficiency through this adjustment. In the later sections of the paper, we investigate several alternate and popular protocols, explored in contemporary literature, to optimize both efficiency and power, wherein we alter both maximum and minimum values of the magnetic field using numerical optimization tools. In the last part we consider more general problem of optimal protocols. The organization of the paper is as follows. Sec. \ref{section:model_and_dynamics}, introduces our setup, presenting the Master equation that governs the dynamics of the spin probabilities, and describes our approach for computing essential thermodynamic quantities. Using the Glauber dynamics, we derived the general equation for magnetization in terms of the general external magnetic field protocol $h(t)$ for the engine model in subsec.~\ref{sec:Analytics_Method}. In subsec. \ref{section:Normal_protocol_long cycle time limit}, we derive precise analytical formulas for work, heat, and efficiency. Subsec. \ref{section:simulation_analytical_results_normal_protocol}, presents a detailed study of numerical and analytical results, examining the behavior of analytically calculated thermodynamic quantities in long cycle time limit and numerical calculations for finite cycle times. In Sec. \ref{sec:optimization_and_optimal_protocol}, we discuss the optimization process and explore various other protocols to achieve maximum efficiency or power by utilizing the gradient descent algorithm. Alternatively, we utilize a completely different optimization technique, using a trial protocol and the variational principle in Sec. \ref{variational} to find the evolution equation for the shape of the entire protocol within a finite cycle time and maximize power and efficiency using both the soft and hard boundary conditions. This approach provides an iterative algorithm that does not require fitting any function and compares power and efficiency with the gradient descent method. Finally, we compare the results obtained from different methodologies in Sec.  \ref{sec:Comparison_all_protocol} and conclude in section \ref{section:Conclusion_futurework}.
 
\section{Definition of Model and Dynamics}
\label{section:model_and_dynamics}
\begin{figure}[!t]
\begin{center}
\includegraphics[width=9cm,height=9cm,angle=0]{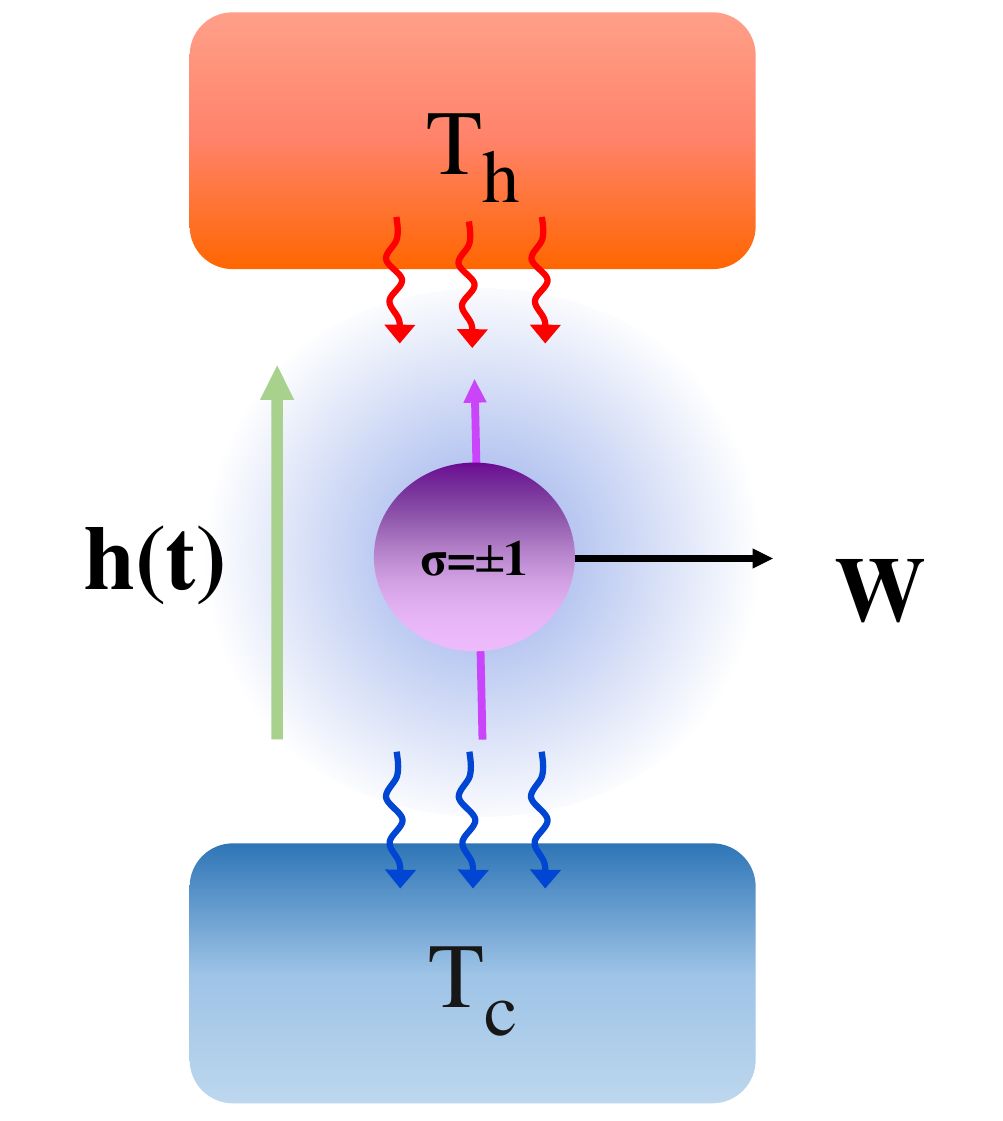}
\caption{A schematic diagram of a stochastic heat engine consisting of a spin under time-dependent magnetic field $h(t)$.}
\label{schematic}
\end{center}
\end{figure}

Our system comprises a single spin, with a magnetic moment $\mu$, driven by an external magnetic field which changes with time denoted by $h(t)$. This is shown in Fig. \ref{schematic}. The Hamiltonian of the system is given as,
\begin{equation}
H = -\mu\sigma h(t),
\label{Hamiltonian}
\end{equation}
where $\sigma=\pm 1$, this can take only two values. Here $+1$ refers to the spin up($\uparrow$) state and $-1$ refers to the spin down($\downarrow$) state. The spin interacts with the connected heat bath at temperatures $T$ (inverse temperature $\beta$, we work in the units where the Boltzmann's constant $k_B=1$).  The spin may flip due to the interaction with the heat bath, this dynamics of the spin is assumed to follow the Glauber dynamics that provides us with the time evolution of the spin \cite{Glauber}. The spin tries to align in the direction of the external driving, but due to the interaction with the heat bath, the spin may flip spontaneously against the direction of the field by drawing energy from the heat bath \cite{Marathe05}. We now derive the time evolution equations for probabilities of different spin configurations, namely $P_{\sigma}(t)$, which gives the probability that at time $t$, the spin is in the state $\sigma=\pm 1$. The Glauber rate of flipping is given by \cite{Glauber},
\begin{eqnarray}
\label{eq:Gluber_rate}
    r_\sigma=r\left(1-\sigma~\tanh{\left[\frac{\mu h(t)}{k_B T}\right]}\right).
\end{eqnarray}
In the continuum-time limit, we define the dimensionless time $t$, which is re-scaled by multiplying it with the rate $r$ in the above equation. We choose the value of $r=0.5$ throughout the paper without the loss of generality. The inverse $r$ also sets a time scale in our problem.

The Master equation for the spin distribution becomes \cite{Marathe05}, 
\begin{eqnarray}
\frac{\partial \hat{P}}{\partial t} = \hat{T}\hat{P},
\end{eqnarray}
where $\mathcal{T}$ is the transition matrix given by, 
\begin{eqnarray}
    \mathcal{T}=\begin{pmatrix}
{\frac{e^{-\beta\mu h(t)}}{Z(t)}}&{-\frac{e^{\beta\mu h(t)}}{Z(t)}}\\ 
{-\frac{e^{-\beta\mu h(t)}}{Z(t)}}&{\frac{e^{\beta\mu h(t)}}{Z(t)}}\\
\end{pmatrix}
\text{and}~~
    \hat{P}=\begin{pmatrix} P_{\uparrow}(t) \\ P_{\downarrow}(t)  \end{pmatrix}.
\end{eqnarray}
Here $Z(t)$ is the instantaneous Canonical partition function of the spin. From the above equation, we can write,
\begin{eqnarray}
    \frac{\partial P_{\uparrow}(t)}{\partial t}&=&-\left(\frac{e^{-\beta\mu h(t)}}{Z}P_{\uparrow}(t)-\frac{e^{\beta\mu h(t)}}{Z}P_{\downarrow}(t)\right)\nonumber\\
\frac{\partial P_{\downarrow}(t)}{\partial t}&=& -\left(-\frac{e^{-\beta\mu h(t)}}{Z}P_{\uparrow}(t)+\frac{e^{\beta\mu h(t)}}{Z}P_{\downarrow}(t)\right)
\end{eqnarray}
$  P_{\uparrow}(t)+P_{\downarrow}(t)=1$ and $P_{\downarrow}(t)=1-P_{\uparrow}(t)$. We now define $P_{\uparrow}(t)=P(t)$ and using above equation we get
\begin{eqnarray}
\dot{P}(t)=-P(t)+\frac{1}{1+e^{-2\beta\mu h(t)}},\nonumber
\end{eqnarray}
that has a formal solution
\begin{eqnarray}
P(t)=P(0) e^{-t}+\int_{0}^{t}\frac{e^{-(t-t')}}{1+e^{-2\beta\mu h(t)}}~dt',
\end{eqnarray}
where $P(0)$ is the initial distribution. We define the average magnetization as, $m(t)=\langle \sigma(t)\rangle=2 P(t)-1$, which satisfies the equation below,
\begin{equation}
\dot{m}(t)=-m(t)+\tanh{\left(\beta\mu h(t)\right)}
\end{equation}
The time evolution of magnetization is given by,
\begin{equation}
m(t)=m(0) e^{-t}+\int_{0}^{t}e^{-(t-t')}\tanh{(\beta\mu h(t'))}dt'
\label{eq:magnetization_eqn}
\end{equation}
The average of internal energy of the system reads,
\begin{equation}
U =-\mu h(t)\sum_{\sigma=\pm 1}\sigma P_{\sigma}(t)
\label{eq:energy}
\end{equation}
After taking the derivative with respect to time and a few simplifications, it gives,
\begin{align}
\dot{U}(t)
=&-\mu\dot{h}(t)\sum_{\sigma=\pm 1}\sigma P_{\sigma}(t)-\mu {h(t)}\sum_{\sigma=\pm 1}\sigma \dot{P}_{\sigma}(t)
= -\mu \dot{h}(t)\left[P_{\uparrow}(t)-P_{\downarrow}(t)\right]-\mu h(t)[\dot{P}_{\uparrow}(t)-\dot{P}_{\downarrow}(t)]\nonumber\\
=&-\mu m(t)\dot{h}(t)-\mu \dot{m}(t)h(t),~~ dU=dW+dQ
\label{firstlaw1}
\end{align}
From the above Eq. (\ref{firstlaw1}), we recognize the average work done as $dW(t)= -\mu \dot{h}(t)m(t)dt$ and the average heat current flowing from the bath to the system as $dQ(t)= -\mu \dot{m}(t)h(t)dt$. After integration, it provides us with the first law of thermodynamics. We can also calculate the heat current independently by considering the energy change $\Delta E $ associated with the spin-flip, namely $\dot{Q}(t)= \sum_{\sigma=\pm 1} r_{\sigma} ~\Delta E~\sigma  ~P_{\sigma}(t)$. When the spin changes from state up to down $(\uparrow \rightarrow \downarrow)$, the value of  $\Delta E$ is  $2\mu h(t)$, and the value is $-2\mu h(t)$ when the spin changes from state down to up $(\downarrow\rightarrow\uparrow)$. 

\subsection{Analytical calculation of the Thermodynamic quantities for an Engine model}
\label{sec:Analytics_Method}
 In this section, we discuss and analyze our system as a heat engine. In order for the system to work like an engine, the spin is our working substance, and the magnetic field is our external parameter, which we change with time to extract work. To achieve this, we construct a cyclic process that undergoes the following set of steps. In the isothermal expansion process, we decrease the strength of the external magnetic field $h(t)$ from $h_{max}$ with time while keeping the bath temperature $T_h$ constant for time $\tau_1 \leqslant t <\tau_2$, the minimum value of the magnetic field at this time is $h_{min}$. During the adiabatic process (instantaneous), there is no heat exchange when we change the temperature from $T_h$ to temperature $T_c$ at time $t=\tau_2$ while the magnetic field is held fixed at its minimum value. The work done in this process is just the change in the internal energy of the system. Then the magnetic field is increased with time from $h_{min}$ while the temperature is held fixed at $T_c$ (this mimics the isothermal compression step) during time $\tau_2\leqslant t<\tau_3$ until it reaches the original field value $h_{max}$. Then, we complete the cycle by mimicking the adiabatic change by adjusting the temperature from $T_c$ to $T_h$ at $\tau_3$. This step also contributes to work given by the change in the internal energy. During two isothermal changes, the total extracted work, $W_1+W_2=W < 0$. Hence we define the heat absorbed from the hot bath as $Q_1>0$ and the heat released to the cold bath as $Q_2<0$. During the adiabatic step, there is no heat exchange between the system and the reservoir, so the work is simply the change of internal energy. 

 The mathematical expressions of these quantities are given in the earlier model section. We can write the time evolution equation of magnetization for each duration following Eq. \eqref{eq:magnetization_eqn}. During the cycle time, $\tau_1 \leqslant t <\tau_2$, and temperature $T_h$ we have
\begin{eqnarray}
    m(t)&=& m(\tau_1) e^{-(t-\tau_1)}+\int_{\tau_1}^{t}e^{-(t-t')}\tanh{(\beta_h\mu h(t'))}dt'\nonumber\\
    &=& m(\tau_1) e^{-(t-\tau_1)}~+~I_{1}(t).
\end{eqnarray}
During the cycle time, $\tau_2 \leqslant t <\tau_3$, temperature $T_c$, the expression of magnetization reads,
\begin{eqnarray}
    m(t)&=& =m(\tau_2) e^{-(t-\tau_2)}+\int_{\tau_2}^{t}e^{-(t-t')}\tanh{(\beta_c\mu h(t'))}dt'\nonumber\\
    &=& m(\tau_2) e^{-(t-\tau_2)}~+~I_{2}(t)
\end{eqnarray}
Using the properties of periodicity  of the protocol, we assume the magnetization is the same at the beginning and end of the protocol ($m(\tau_1)=m(\tau_3)$ also at $\tau_2$  the magnetization is continuous ( $m(\tau_2^-)=m(\tau_2^+)=m(\tau_2)$) 
We calculate the following quantities,
\begin{equation}
m(\tau_1)=\dfrac{I_1(\tau_2)e^{-(\tau_3-\tau_2)}+I_2(\tau_3)}{1-e^{-(\tau_3-\tau_1)}}, ~~  
m(\tau_2)=m(\tau_1)e^{-(\tau_2-\tau_1)}+I_1(\tau_2)
\label{sigmaRTs}
\end{equation}
Using the definition of work done given at the end of section \ref{section:model_and_dynamics}. We can write the work done during $\tau_1 \leqslant t <\tau_2$ as,
\begin{eqnarray}
W_{1}(t)&=&- \mu m(\tau_1)\int_{\tau_1}^t\dot{h}(t')  e^{-t'}dt'~-~\int_{\tau_1}^t \int_{\tau_1}^{t}e^{-(t'-t'')}\dot{h}(t')\tanh{(\beta_h\mu h(t''))}dt'  dt''\nonumber\\
%&=& -\frac{2\mu m(0)h_{0}(N-1)}{N\tau}\left(e^{-t}-1\right)~-~\int_0^t \int_{0}^{t}e^{-(t'-t'')}\dot{h}(t')\tanh{(\beta_h\mu h(t''))}dt'  dt''
\label{eq:work_first}
\end{eqnarray}
Similarly, for the duration $\tau_2 \leqslant t <\tau_3$ the work done is, ,
\begin{eqnarray}
W_{2}(t)&=& - \mu m(\tau_2) \int_{\tau_2}^t\dot{h}(t')  e^{-(t'-\tau_2)}dt'~-~\int_{\tau_2}^t \int_{\tau_2}^{t}e^{-(t'-t'')}\dot{h}(t')\tanh{(\beta_c\mu h(t''))}dt'  dt''\nonumber\\
%&=&-\frac{2\mu m(0)h_{0}(N-1)}{N\tau}\left(1-e^{-(t-\tau/2)}\right)~-~\int_{\tau/2}^t \int_{\tau/2}^{t}e^{-(t'-t'')}\dot{h}(t')\tanh{(\beta_c\mu h(t''))}dt'  dt''\nonumber\\
\label{eq:work_second}
\end{eqnarray}
During the adiabatic step, the work is
\begin{equation}
W_3=-\mu m(\tau_2)(h(\tau_2^+)-h(\tau_2^-)),~~ W_4=-\mu m(\tau_3)(h(\tau_3^+)-h(\tau_3^-)),   
\label{adiabatic_work_1}
\end{equation}
Similarly, the heat for each duration of isothermal steps,
\begin{eqnarray}
Q_{1}(t)&=& \mu\int_{\tau_1}^t h(t')m(t') dt'-\mu\int_{\tau_1}^t  h(t')\tanh{\left(\beta_h\mu h(t')\right)}dt'
\label{Heat_first}
\end{eqnarray}
\begin{eqnarray}
Q_{2}(t)&=&\mu\int_{\tau_2}^t h(t')m(t') dt'-\mu\int_{\tau_2}^t  h(t')\tanh{\left(\beta_c\mu h(t')\right)}dt'
\label{Heat_second}
\end{eqnarray}
We define the efficiency of the engine as the ratio of total work ($W=W_1+W_2+W_3+W_4$) extracted from the system and the heat ($Q_1$) flow from the hot bath,
\begin{eqnarray}
  \eta &=&-\frac{W}{Q_1} \nonumber\\
  \label{eq:effi_all}
\end{eqnarray}
Also, we define several important thermodynamic quantities like average power and average entropy production as,
\beqa
~~P = \frac{|W_1~+~W_2+W_3+W_4|}{\tau}, 
~~ S =-\dfrac{Q_1}{T_h}-\dfrac{Q_2}{T_c}, \nonumber\\
\label{eq:power_entropy_all}
\eeqa 

\subsubsection{Calculation in the long cycle time limit}
\label{section:Normal_protocol_long cycle time limit}
For the long cycle time limit, the spin configuration probability $P_{\sigma}(t)$ can be expressed in terms of Canonical distribution, 
\begin{eqnarray}
P_{\sigma}(t)=\frac{e^{-\beta\mu h(t)\sigma}}{(e^{\beta\mu h(t)}+e^{-\beta\mu h(t)})}=
\frac{e^{-\beta\mu h(t)\sigma}}{2 \cosh{(\beta\mu h(t))}}
 \label{probability}
\end{eqnarray}
From the above equation, we calculate the magnetization, 
\begin{eqnarray}
m(t) =\langle \sigma(t)\rangle= \tanh{\left(\beta\mu h(t)\right)}
\end{eqnarray}
We calculate all the thermodynamic quantities for a general protocol $h(t)$ for the long cycle time limit.
The work in an isothermal step reads
\begin{eqnarray}
\label{eq:general_h(t)_work}
W_{iso}=-(1/\beta)\int_{t'}^{t''} \beta\mu\dot{h}(t) \tanh(\mu\beta h(t))dt=-(1/\beta) \log\left[\dfrac{\cosh(\mu\beta h(t''))}{\cosh(\mu\beta h(t'))}\right]
\end{eqnarray}
 For the adiabatic jump $dQ=0$, during this stroke, the work done is calculated by evaluating the change of internal energy of the system. 
The heat exchanges during isothermal transitions reads,
\begin{eqnarray}
\label{eq:general_heat_calculation}
Q=\int_{t'}^{t''} \mu h(t)\dfrac{d}{dt}(\tanh(\mu\beta h(t)))dt=-\mu h(t)\tanh(\mu\beta h(t)) \bigg\rvert_{t'}^{t''}~+~(1/\beta) \log\left[\dfrac{\cosh(\mu\beta h(t''))}{\cosh(\mu\beta h(t'))}\right]\nonumber\\
\end{eqnarray}

The adiabatic work is,
\begin{equation}
W_{adia}=-\mu \tanh(\mu\beta h(t'))(h(t'')-h(t')),   
\label{adiabatic_work_large_cycle}
\end{equation}
In the above Eqs. \eqref{eq:general_h(t)_work}, \eqref{eq:general_heat_calculation},      \eqref{adiabatic_work_large_cycle}, $t'$ and $t''$ are the initial and final time of individual stroke (for both isothermal and adiabatic process). From the above equation, we conclude that the work done and heat for a long cycle time limit depend only on the initial and final values of the protocol $h(t)$. Thus becoming independent of the form of the protocol.
 
\begin{figure}[!t]
\begin{center}
\includegraphics[width=8cm,height=6cm,angle=0]{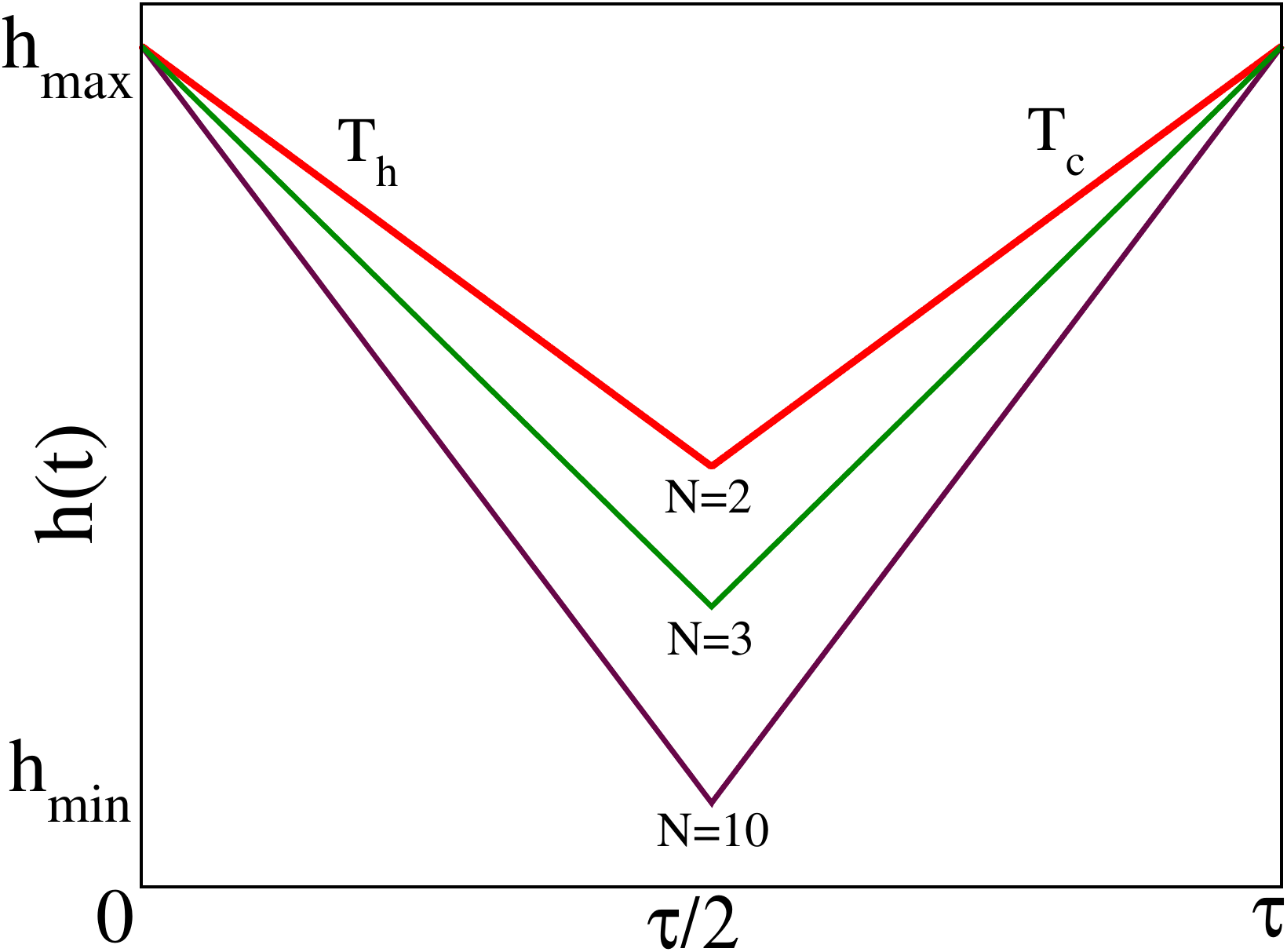}
\caption{A particular type of the time-dependent protocol of the stiffness $h(t)$. $h_{max}$ and $h_{min}$ are the maximum and minimum values of the field. }
\label{protocolfig}
\end{center}
\end{figure}

Now, in Fig. \ref{protocolfig}, we discuss a particular protocol that mimics the steps of an engine model. Here, the cycle is constructed where $\tau_1=0$, $\tau_2=\tau/2$ and $\tau_3=\tau$ and we also introduce the maximum value of the magnetic field as $h_{max}$ and $h_{min}$. $N$ is a parameter that can take any value. By changing $N$, we choose any minimum value of the external field, and further, we  study if the efficiency depends on the ratio of $\frac{h_{max}}{h_{min}}=N$ 
\begin{equation}
h(t)= 
\begin{cases}
 (h_{min}-h_{max})\dfrac{2t}{\tau}+h_{max} & , 0\leq t < \tau/2,\\
(h_{max}-h_{min})\dfrac{2t}{\tau}+2h_{min}-h_{max}& , \tau/2\leq t <\tau,
\end{cases}
\label{protocol}
\end{equation}

Using Eq. \eqref{eq:general_h(t)_work}, the work is calculated for both the isothermal steps as,
\begin{equation}
\label{eq:work1,2_quasi}
W_1=-k_{B}T_h\log{\left[\frac{\cosh{(\mu\beta_h h_{min})}}{\cosh{(\mu\beta_hh_{max})}}\right]},\quad 
 W_2=-k_{B}T_c\log{\left[\frac{\cosh{(\mu\beta_c h_{max})}}{\cosh{\left(\mu\beta_c h_{min}\right)}}\right]},
\end{equation}
The adiabatic work is ,
\begin{equation}
W_3=-\mu \tanh(\mu\beta_h h_{min})(h_{min}-h_{min})=0,~~ W_4=-\mu \tanh(\mu\beta_c h_{max})(h_{max}-h_{max})=0,   
\label{adiabatic_work_quasi}
\end{equation}
So the total work for this particular protocol reads, $W=W_1+W_2+W_3+W_4=W_1+W_2$, where, $\beta_h=\frac{1}{k_{B}T_h}$ and  $\beta_c=\frac{1}{k_{B}T_c}$ are the inverse temperatures and  $T_h$ and $T_c$ are the hot and cold reservoir temperatures. $Q_1$ is the average heat that originates from the hot reservoir or left bath, and $Q_2$ is the heat released to the cold reservoir or right bath. We calculate heat using Eq. \eqref{eq:general_heat_calculation}, which reads,
\begin{eqnarray}
Q_1 &=& \mu h_{max}\tanh{(\mu\beta_c h_{max})}-\mu h_{min}\tanh{\left(\mu \beta_h h_{min}\right)}~+~k_{B}T_h\log{\left[\frac{\cosh{(\mu\beta_h h_{min})}}{\cosh{(\mu\beta_hh_{max})}}\right]}\nonumber\\
 \label{eq:heat1_quasi}
\end{eqnarray}
\begin{eqnarray}
Q_2&=& \mu h_{min} \tanh{\left(\mu\beta_h h_{min}\right)}-\mu h_{max} \tanh{(\mu\beta_c h_{max})}~+~k_{B}T_c\log{\left[\frac{\cosh{(\mu\beta_c h_{max})}}{\cosh{\left(\mu\beta_c h_{min}\right)}}\right]}
\label{eq:heat2_quasi}
\end{eqnarray}
 The efficiency of the engine is calculated using Eq. \eqref{eq:effi_all}, and the expression reads,
\begin{eqnarray}
\eta&=&\dfrac{k_{B}T_h\log{\left[\frac{\cosh{(\mu\beta_h h_{min})}}{\cosh{(\mu\beta_hh_{max})}}\right]}+k_{B}T_c\log{\left[\frac{\cosh{(\mu\beta_c h_{max})}}{\cosh{\left(\mu\beta_c h_{min}\right)}}\right]}}{\mu h_{max}\tanh{(\mu\beta_c h_{max})}-\mu h_{min}\tanh{\left(\mu \beta_h h_{min}\right)}~+~k_{B}T_h\log{\left[\frac{\cosh{(\mu\beta_h h_{min})}}{\cosh{(\mu\beta_hh_{max})}}\right]}}
  \label{eq:effi_quasi}
\end{eqnarray}
Similarly, we can calculate the power and the entropy using Eq. \eqref{eq:power_entropy_all}.
\subsection{ Results from simulation and comparison with analytical expression}
\label{section:simulation_analytical_results_normal_protocol}
In this section, we analyze and compare the analytical and numerical results for a particular protocol already discussed in ~Fig. \ref{protocolfig}. We fix the number of steps to complete the cycle of duration $\tau$ to $\sim 10^4$ the time step $dt\sim \tau/10^4$. We work in the units where Boltzmann constant $k_B=1$ and the magnetic moment $\mu=1$. To simulate the system, we connect the spin to a heat bath at temperature $T$, and its flip dynamics are governed by the Glauber rates ($r_{\sigma}$) given in Eq. \eqref{eq:Gluber_rate}.  We choose our rate constant $r$ to be $0.5$. The initial value of the spin is chosen randomly, and the system is run through several cycles of duration $\tau$ till it reaches the steady state. All the average quantities like magnetization, work, and heat are calculated in the steady state. In the steady state, in a given time step, if the spin flips, we evaluate the heat exchanged between the spin and the heat reservoir. Let's assume the magnetic field at $n^{th}$ step is $h_n$, and the spin value is $\sigma_n$. Now, after $dt$ time, the value of the magnetic field changes from  $h_n$ to $h_{n+1}$. During this step, the  work,  $\Delta W=-\mu\sigma_{n}\Delta h_n$,  where $\Delta h_n=h_{n+1}-h_n$.  If initially the spin is in a state of $\sigma_{n}$ and it flips into a state $\sigma_{n+1}$, we calculate the heat as, $\triangle Q=-\mu h_{n+1}(\sigma_{n+1}-\sigma_{n})$. Due to the presence of an external magnetic field, the spin tries to align in the direction of the magnetic field. However, due to the effect of the heat bath at the some temperature $T$, there is a finite probability that the spin may draw some energy from the heat bath and flip against the magnetic field. For a single spin system, we define magnetization as the expectation value of the spin, namely $m(t)=\langle\sigma(t)\rangle$. So, at a constant temperature $T_h$, if the magnetic field changes slowly (than the relaxation time of the system), then the magnetization will follow the change the magnetic field. 

\begin{figure}
 \hspace{-1cm}  
\includegraphics[width=7.5cm,height=6.0cm,angle=0]{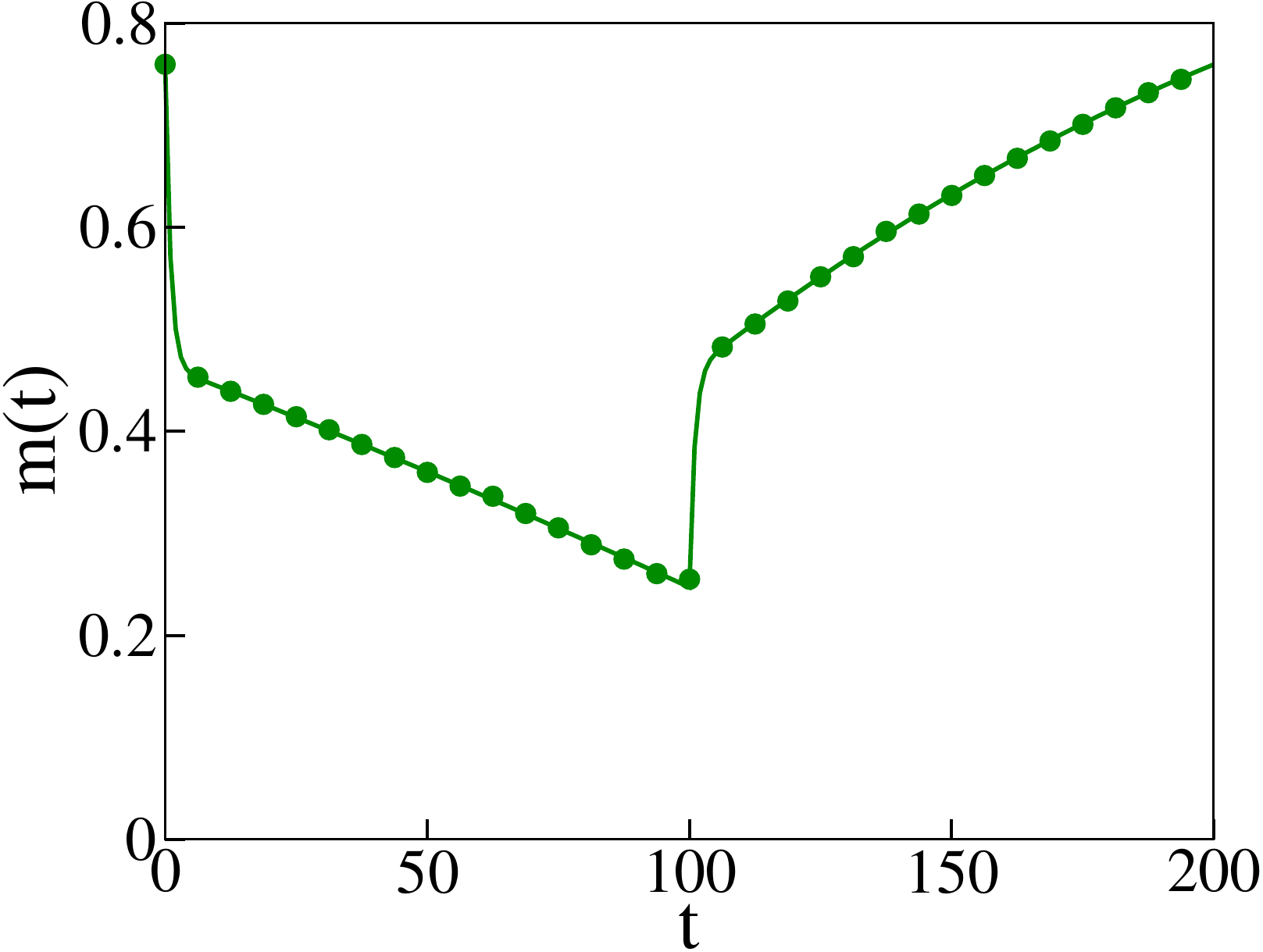}~\\~\\
\includegraphics[width=7.5cm,height=6.0cm,angle=0]{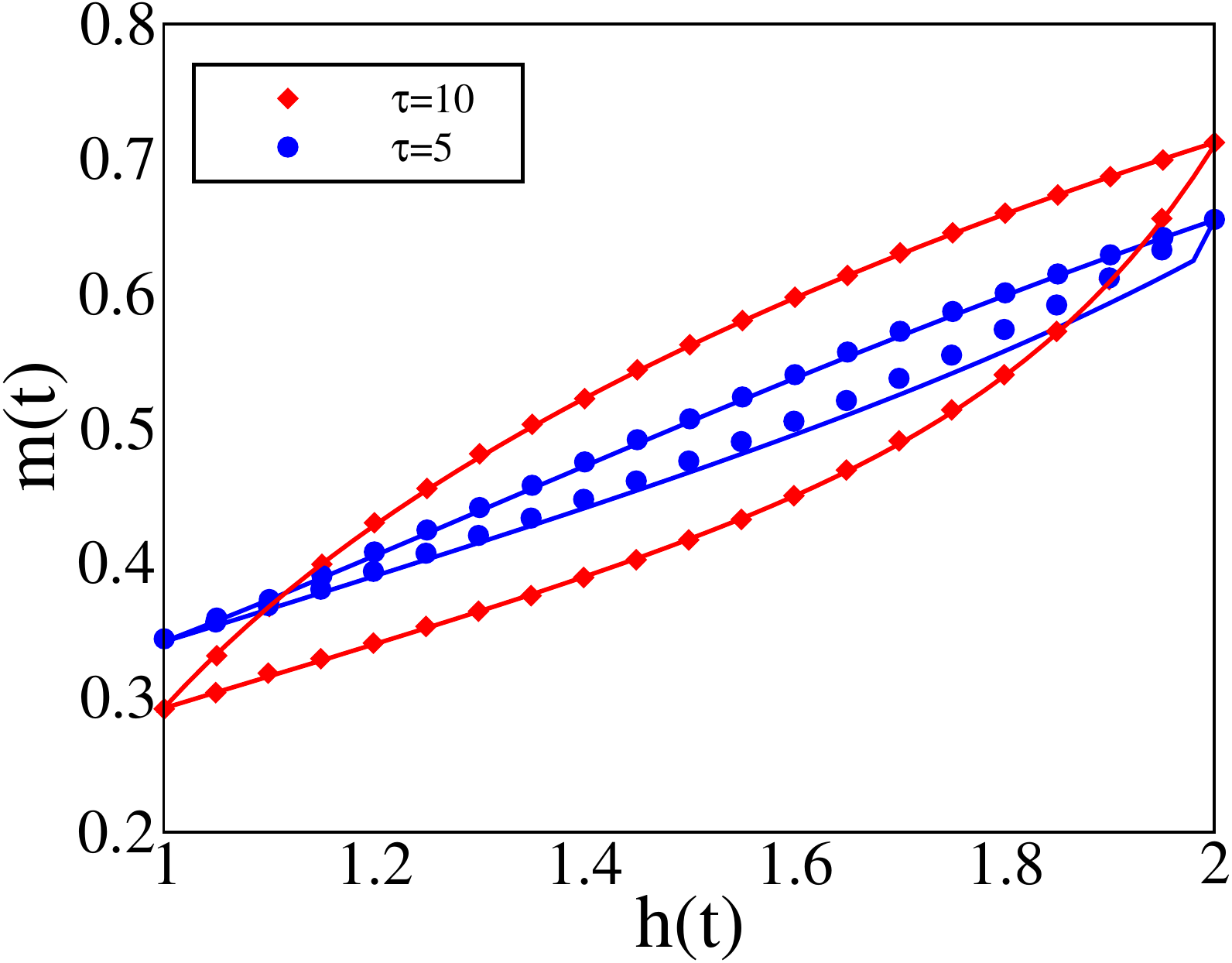}~~~
\includegraphics[width=7.5cm,height=6.0cm,angle=0]{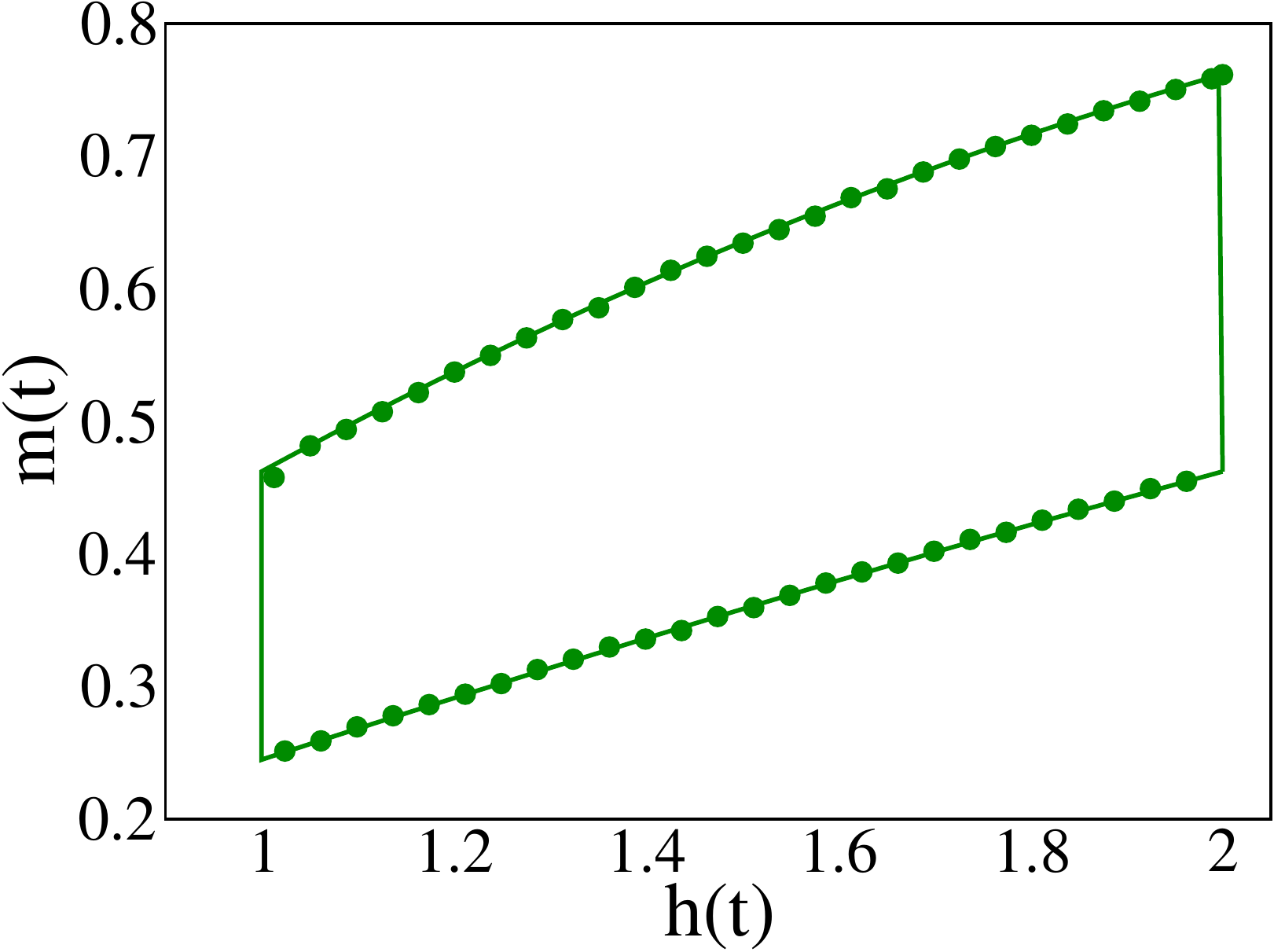}
    \caption{(top) Simulation and numerical results of the  magnetization $m(t)$ with time $t$. (bottom left) The plot of magnetization $m(t) $ with external magnetic field $h(t)$. Simulation and numerical results $m(t) $ with $h(t)$ for different $\tau$ for the non-quasi static case and for quasi-static case (bottom right). The plot mimics the Stirling engine-like cycle for the long cycle time limit. The symbols represent the simulation, and the solid lines represent the numerical/analytical results. Parameters are used $T_{h}=4.0$, $T_{c}=2.0$, $\mu=1.0$, $h_{max}=2.0$, $h_{min}=1.0$, $r=0.5$, $N=2$, $\tau=200$.}
\label{sigma}  
\end{figure}

   In Fig. \ref{sigma} (top), we plot the average magnetization as a function of the time $t$. During the first half of the cycle $0\leq t < \tau/2$, the system operates at a higher and constant temperature ($T_h$), known as the isothermal step. During this step, the magnetization decreases if we decrease the magnetic field. During the adiabatic step,  at $\tau/2$, the value of the external magnetic field is constant, but a sudden temperature switch occurs from $T_h$ to $T_c$, which decreases the fluctuations due to the thermal bath and increases magnetization. During the period $\tau/2\leq t < \tau $, at constant temperature $T_c$, the magnetization continues to rise as the probability of spin alignment with the external magnetic field increases. At the end of the cycle, another temperature switch occurs from lower ($T_c$) to higher ($T_h$), resulting in lower magnetization and the cycle continues. In Fig. \ref{sigma} (bottom left) and (bottom right), we plot the magnetization $m(t)$ with the external magnetic field $h(t)$, which provides some valuable information related to the work and heat involved in the system. The area enclosed by the curve represents the work done by the system. This is similar to the $P-V$ diagram in usual thermodynamic engines. Shows us that in the long cycle time limit, this diagram reaches the Stirling protocol-like cycle. Fig. \ref{sigma} (bottom left) represents the hysteresis loop for a shorter cycle time that looks like usual hysteresis loops without the adiabatic jumps, as seen for the case of the long cycle time discussed before. To note, in both these and the latter diagrams, continuous lines are the numerical or analytical results calculated from the equations described in Sec. \ref{sec:Analytics_Method}, and the symbols represent simulation results. 

\begin{figure}[t!]
\begin{center}
\includegraphics[width=7.2cm,height=5cm,angle=0]{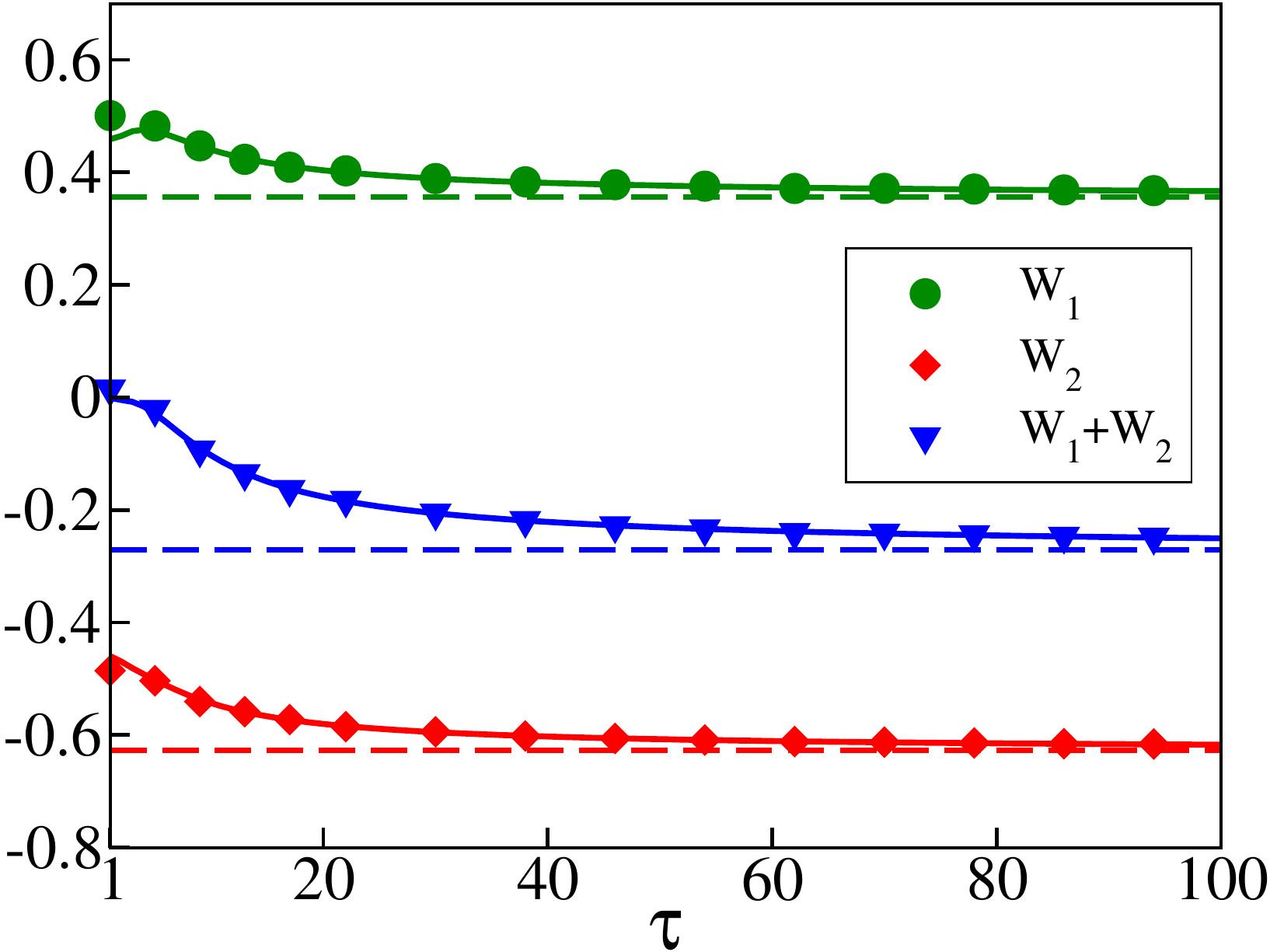}~~~~
\includegraphics[width=7.2cm,height=5cm,angle=0]{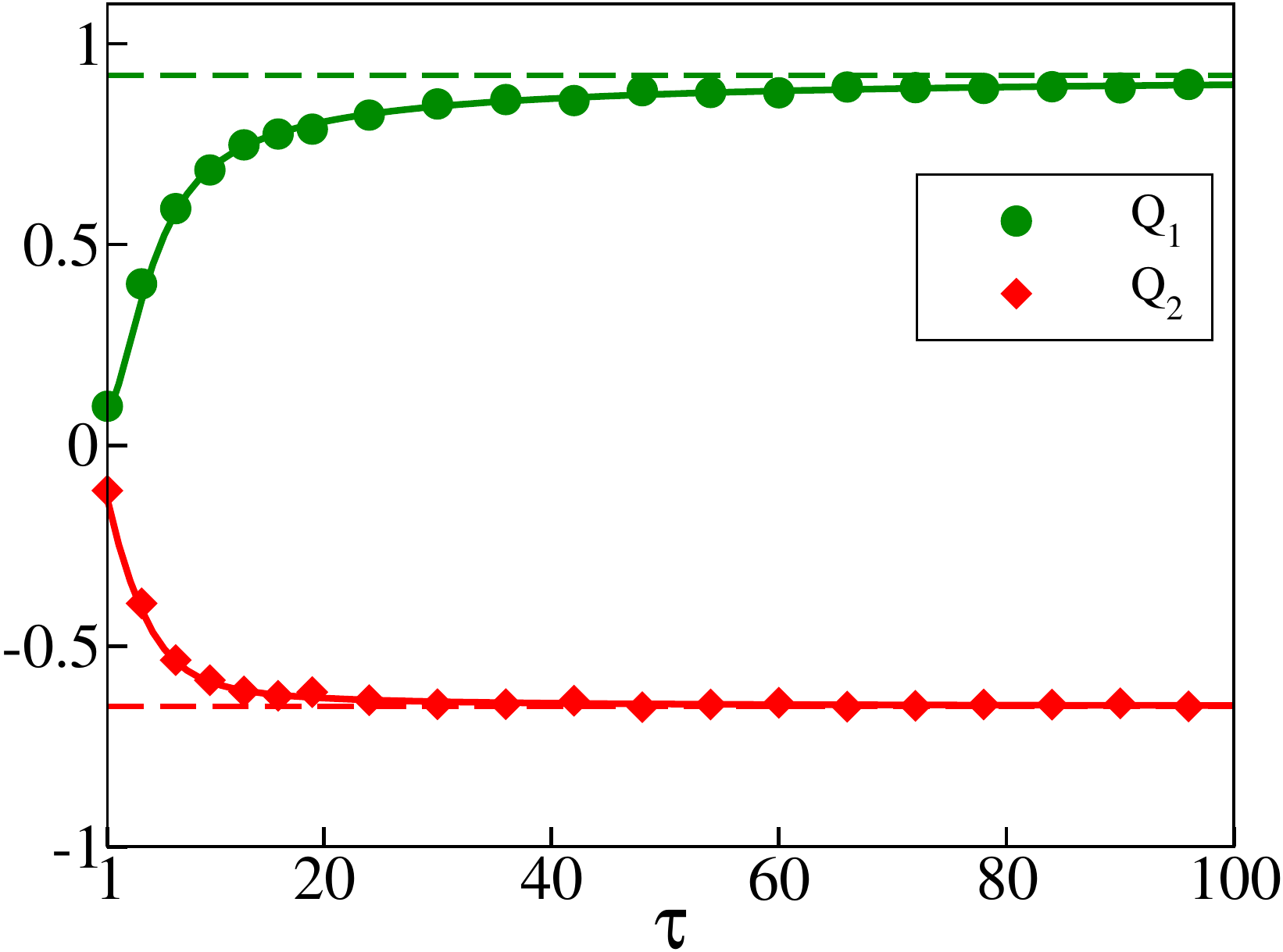}~\\~\\
\includegraphics[width=7.2cm,height=5.2cm,angle=0]{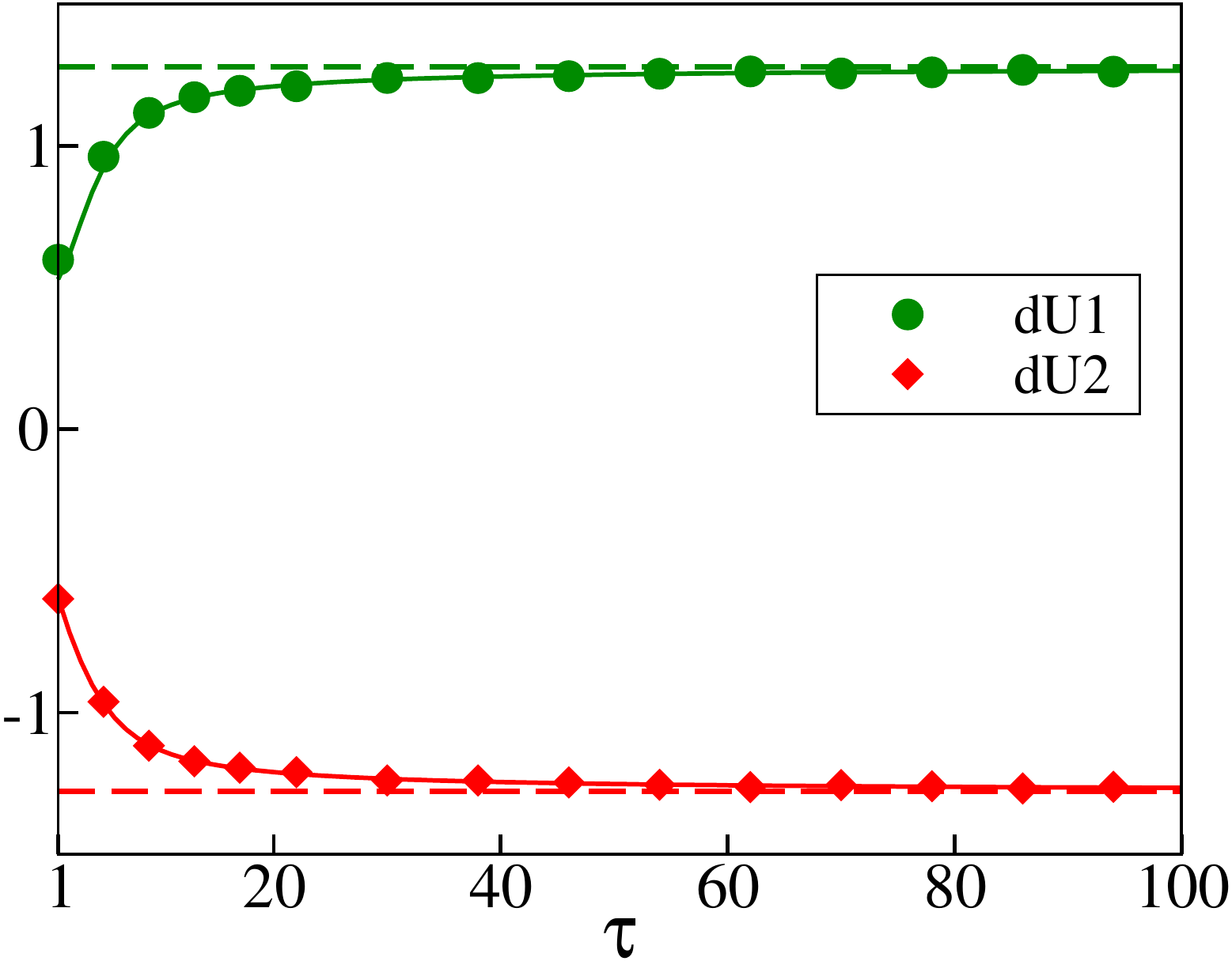}~~
\includegraphics[width=7.2cm,height=5.2cm,angle=0]{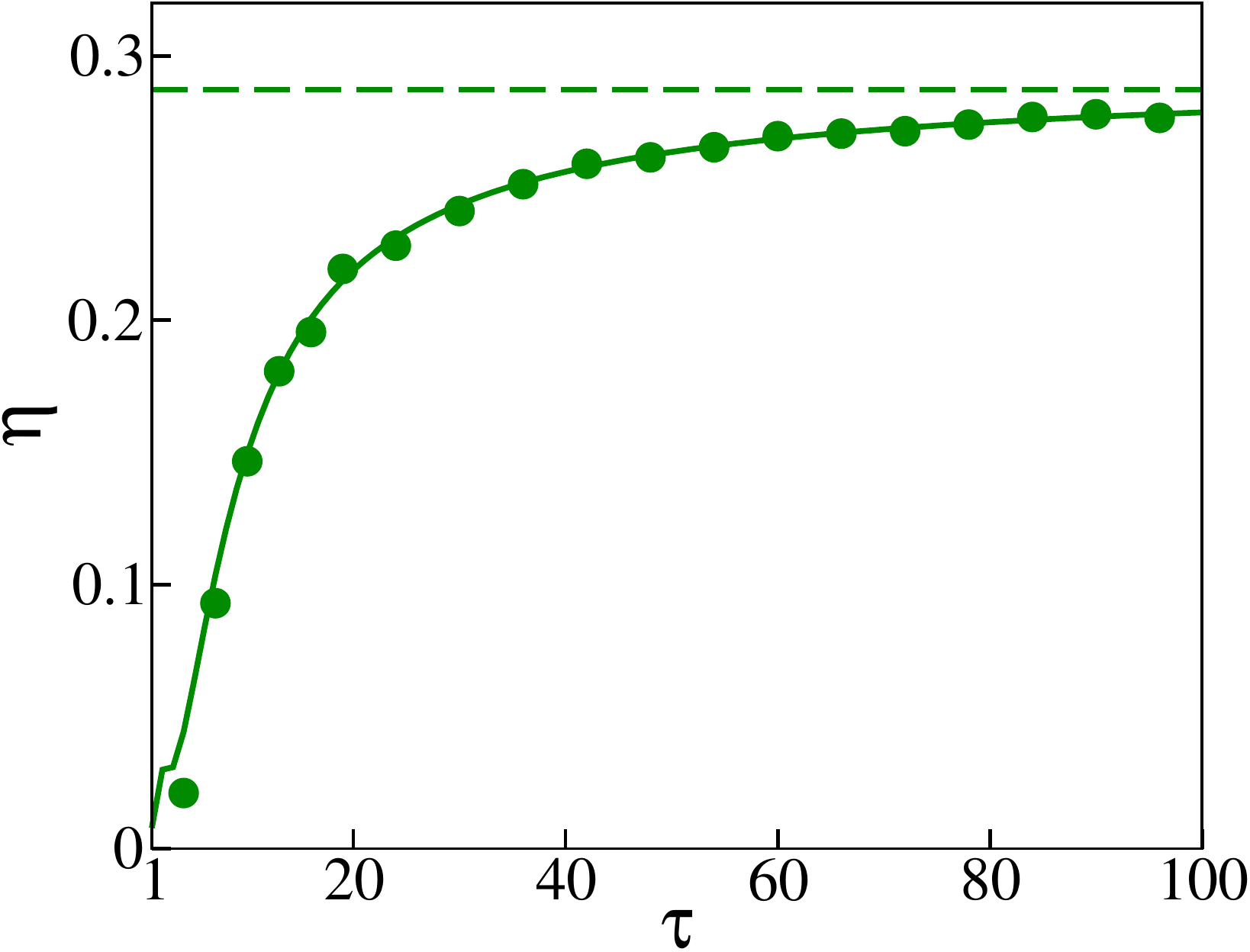}
\caption{(top left) Individual works and total work $ W_{1} $, $ W_{2}$ and $W_{1} + W_{2}$ with cycle time $\tau$. (top right) $ Q_{1} $, $Q_{2}$  heats with cycle time $\tau$. (bottom left) Change of internal energy and total work  $dU_1, dU_2 $ with cycle time $\tau$. (bottom right) $\eta $ efficiency with cycle time $\tau$.  We use the parameters are given below,  $T_{h}=4.0$, $T_{c}=2.0$, $\mu=1.0$, $h_{max}=2.0$, $h_{min}=1.0$, $h_{min}=1.0$, $r=0.5$, $N=2$.  Unless otherwise specified, symbols in all the figures represent simulation results and solid lines of analytical results, and the dashed horizontal lines represent the values at the long cycle time limit.}
\label{workplot}
\end{center}
\end{figure}

In Fig. \ref{workplot}, we plot average accumulated work and heat within a cycle as a function of $\tau$, using data obtained from analytical formulae and simulations. We calculate work analytically using Eqs. \eqref{eq:work_first}, \eqref{eq:work_second},  and the long cycle time limit calculations are done using Eqs. \eqref{eq:work1,2_quasi}. It is evident from the graph that the total work $W_{1} + W_{2}$ initially decreases and then saturates at a particular negative value (in a long cycle time limit). So, the total thermodynamic work along a cycle in a long cycle time limit is negative ($W>0$), suggesting that the model works as an engine. Similarly, in Fig. \ref{workplot} (top right), we plot accumulated heat as a function of $\tau$. We calculate heat analytically using Eq. \eqref{Heat_first}, \eqref{Heat_second}; the long cycle time limit results are obtained using Eq. \eqref{eq:heat1_quasi}-\eqref{eq:heat2_quasi}. $Q_{1}$ is the heat the system absorbs from the hot bath. $Q_{2}$ is the heat release in the cold bath. In any cycle limit, the numerical results are consistent with our analytical predictions made in the previous part.
 In Fig. \ref{workplot}, in the (bottom left), we plot the internal energy $dU_1$ and $dU_2$ as a function of $\tau$, we evaluate analytical results using Eq. \eqref{eq:energy}.
 In the (bottom right), we show the plot of $\eta$ with cycle time $\tau$.  Theoretically, we calculate the efficiency for any cycle time $\tau$. The efficiency increases and saturates to a specific value as we go from the short cycle time to the long cycle time limit. Theoretically, we obtain the efficiency using Eq. \eqref{eq:effi_all} and Eq. \eqref{eq:effi_quasi} for long and short cycle limits accordingly.
\begin{figure}[!t]
\includegraphics[width=7.5cm,height=5.0cm,angle=0]{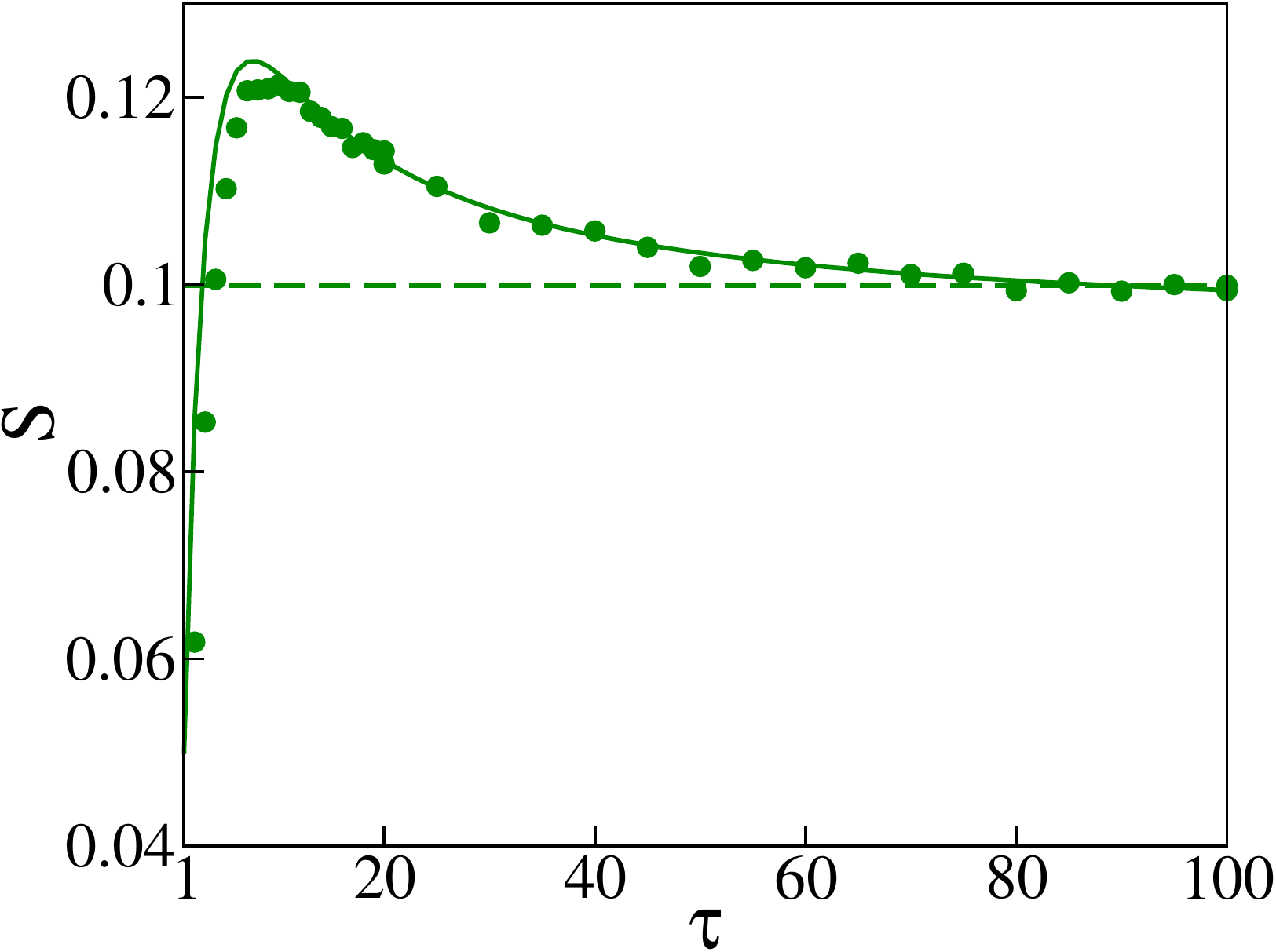}~~
\includegraphics[width=7.5cm,height=5.2cm,angle=0]{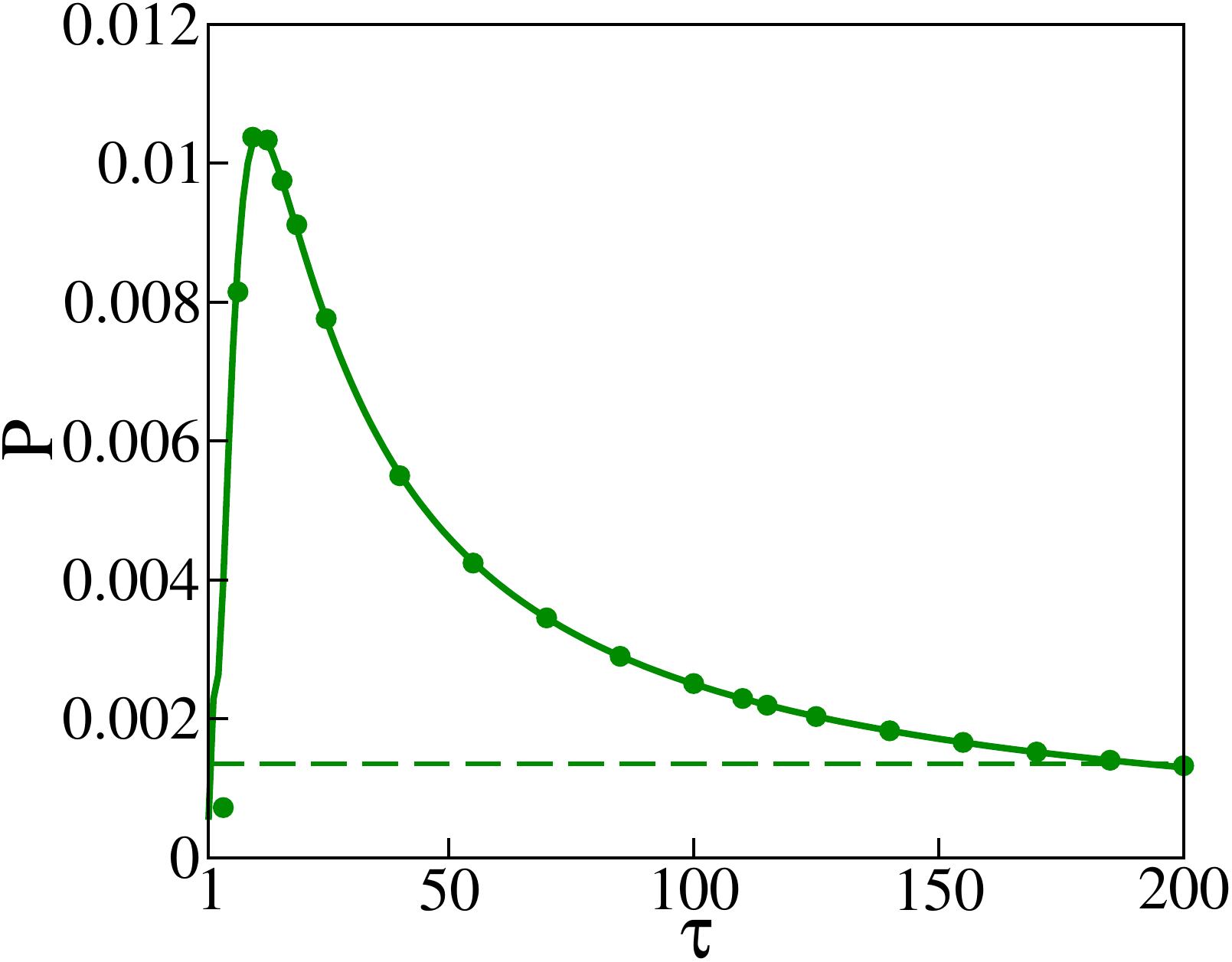}~\\~\\
~\includegraphics[width=7.8cm,height=5cm,angle=0]{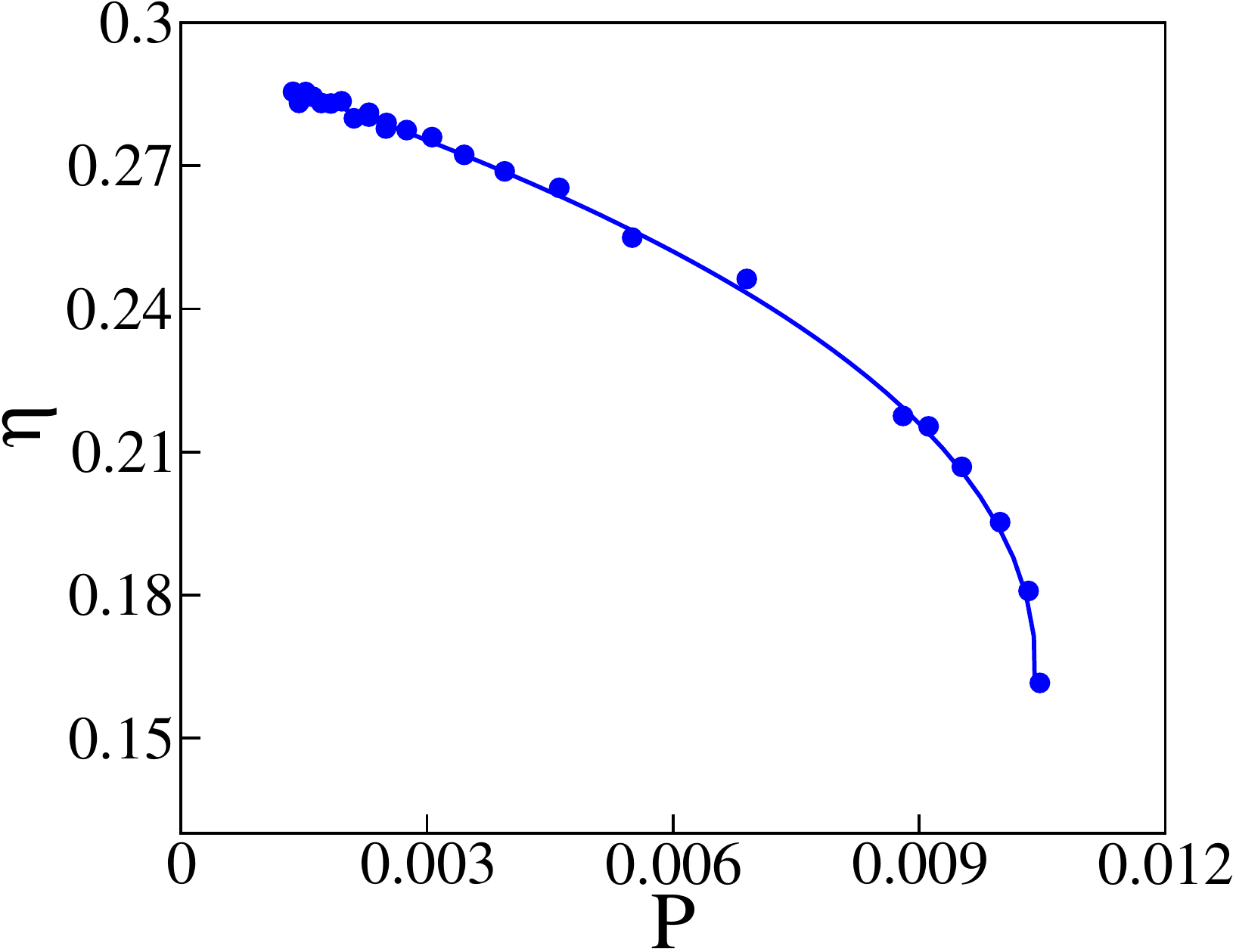}~~~~
\includegraphics[width=7.0cm,height=5cm,angle=0]{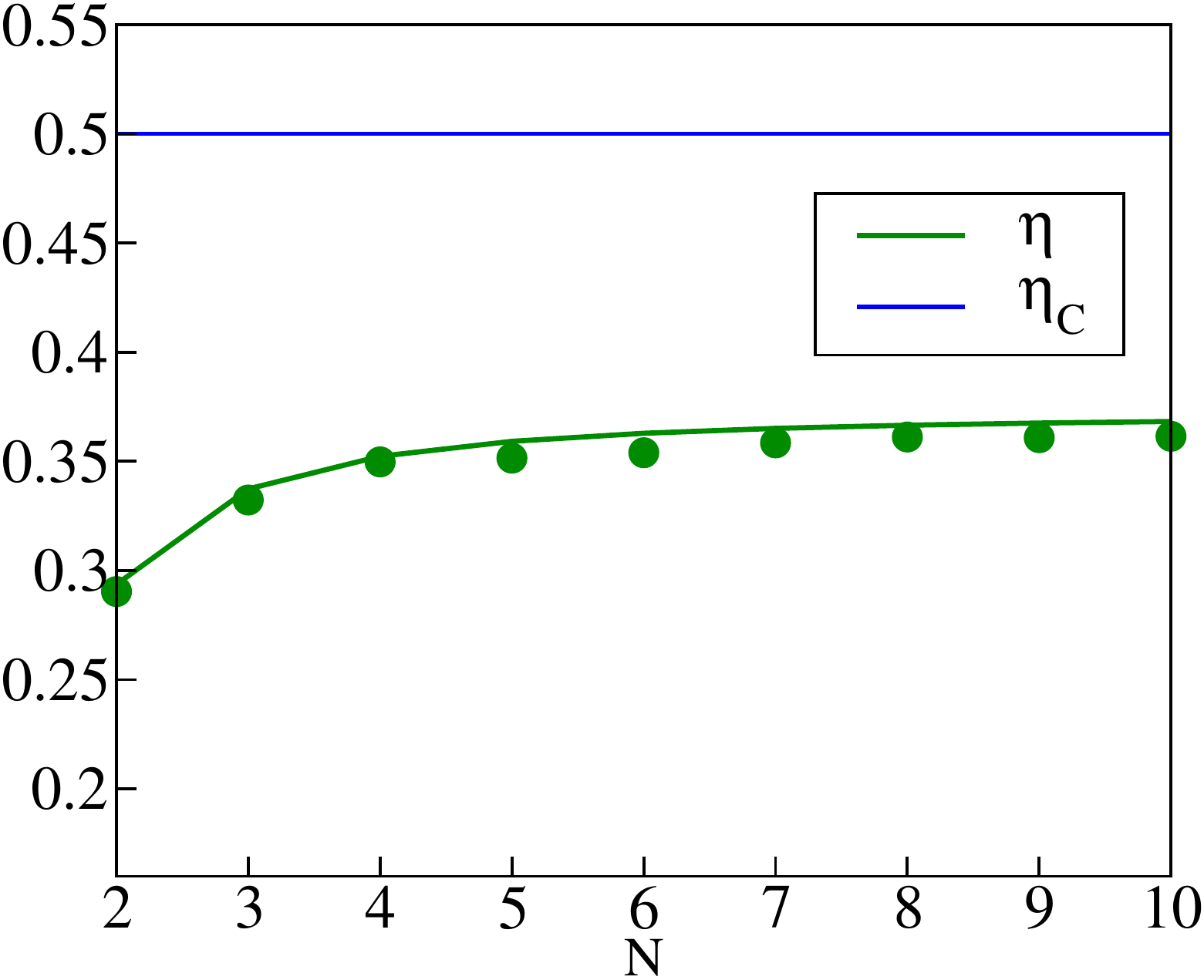}
\caption{Entropy (top left) and power (top right) as a function of the cycle time $\tau$. (bottom left) Represents the Power vs $\eta$. (bottom right) Represents variations of $\eta$ with $N$. Parameters used are  $T_{h}=4.0$, $T_{c}=2.0$, $\mu=1.0$, $h_{max}=2.0$, $r=0.5$ and $N=2$.}
   \label{entr}
\end{figure}
In Fig. \ref{entr}, we see in the (top right) that the entropy $S>0$ is positive, reaches a maximum, and
saturates in the long cycle time limit. We also plot power $P$ defined in Eq. \eqref{eq:power_entropy_all} as a function of the
cycle time. We note that the power is maximum at a value of $\tau$ where the $\eta$  is much lower than the efficiency at the long cycle time limit.  In the average power versus average efficiency plot (see Fig. \ref{entr} (bottom left)), the efficiency decreases as the power increases, indicating the powerful machine has lower efficiency. In the (bottom right), we plot $ \eta $ vs. $N$ where we define $ h_{min}=\frac{h_{max}}{N}$. We want to see if the efficiency increases while changing the minimum value of the magnetic field at $\tau/2$. All the previous results were obtained for $N=2$. In the Fig. \ref{entr} (bottom right), initially, the efficiency is increases, then it saturates at a particular value as $N$ is increased, but as expected, it never matches the Carnot efficiency.

\section{Optimization and Optimal Protocols}
\label{sec:optimization_and_optimal_protocol}
In the earlier sections, Sec. \ref{section:Normal_protocol_long cycle time limit}, Sec. \ref{section:simulation_analytical_results_normal_protocol},  we discussed the results for efficiency, power, and various other thermodynamic properties for a linear protocol. Furthermore, we explored strategies to enhance the engine's efficiency by adjusting the ratio between the minimum and maximum values of the protocol. There are several numerical and theoretical approaches to optimize the engine's efficiency. In the context of a heat engine, optimization refers to achieving the maximum efficiency and maximum power or both simultaneously. By fine-tuning various parameters, one can achieve better engine performance. There is a growing interest in developing engineered techniques and experimental setups, designing engineered swift equilibration (ESE) processes where one can design efficient engines \cite{Plata19, Ignatio16}. One can optimize the system's output under constraints on the state of the working medium \cite{Zhang}.
The parameters we can control experimentally are the temperatures of baths and the intensity of an optical trap for the Brownian heat engine. Under experimentally motivated constraints \cite{Plata19, Abiuso22, Zhong22}, one can aim to determine the optimal protocols for calculating optimal power and efficiency. In the underdamped regime, fixing efficiency and optimizing for power can lead to the protocols taking exponential form \cite{Chen22}. In the realm of the Overdamped Brownian system \cite{Ye22}, a well-studied protocol is the Carnot-Otto cycle, composed of two isotherms and two adiabats. The piece-wise constant protocol provides the maximum efficiency for an engine model and does not depend on the working substance; hence, it is known as the maximum efficiency protocol.   

One can optimize the extracted work and heat for a spin system by fine-tuning the external magnetic field. Here,  we calculate all thermodynamic quantities analytically using the piece-wise constant protocol given in Eq. \eqref{maximum_efficiency_protocol} to achieve maximum efficiency. This is similar to the protocol used in case of the Langevin system discussed above, except the trap stiffness used there is replaced by magnetic field. As detailed below, we follow the standard mathematical approach and derive the essential formulae to calculate the average of work, heat, efficiency, and other various quantities specific to this protocol. 

\begin{equation}
h(t)= 
\begin{cases}
h_{max} & , 0< t \leq \tau/2,\\
h_{min} & , \tau/2< t \leq\tau,
\end{cases}
\label{maximum_efficiency_protocol}
\end{equation}

In the Piecewise constant protocol Eq. \eqref{maximum_efficiency_protocol}, two isothermal processes are connected by two adiabatic processes at $\tau/2$ and $\tau$. The temperature is maintained at $T_h$ in the first half-cycle, and as $T_c$ in the second half-cycle. During each isothermal branch, no work is done ($dW=0$) because the field is constant, but heat is transferred from the heat bath to the system and can simply be expressed in terms of the internal energy change. For the Adiabatic process, there is no heat exchange ($dQ=0$), so the work done is simply presented as the change of internal energy of the system. Thus, we calculate the work during the adiabatic process followed by Eq. \eqref{firstlaw1} (see Sec. \ref{section:model_and_dynamics}),
\begin{eqnarray}
  W_1=dU_1=\mu m(\tau/2)(h_{max}-h_{min}), ~~ W_2=dU_2=\mu m(\tau)(h_{min}-h_{max}) 
\end{eqnarray}
The total output work  and input heat during the first Isothermal process is again calculated using Eq. \eqref{firstlaw1},
\begin{eqnarray}
\label{eq:PWCW}
 W=\mu (h_{max}-h_{min})(m(\tau/2)-m(\tau))
= -\mu \Delta h \Delta m ,~~ Q_1=\mu h_{max}(m(\tau)-m(\tau/2))
= \mu h_{max}\Delta m   \nonumber\\
\end{eqnarray}
So efficiency reads, 
\begin{equation}
\label{eq:PWCeffi}
\eta=\frac{\mu \Delta h \Delta m}{\mu h_{max}\Delta m}
= \frac{h_{max}-h_{min}}{h_{max}}= 1-\frac{h_{min}}{h_{max}}
\end{equation}
it shows that the efficiency is independent of the system dynamics as it only depends on the ratio between the minimum and maximum value of the external control parameter. This cycle configuration provides the maximum achievable efficiency for a given system when the ratio~ $\dfrac{h_{min}}{h_{max}}$~ is at its minimum value as seen from Eq. \eqref{eq:PWCeffi}. The value of efficiency is always bounded by the Carnot efficiency ($\eta_C$), where ~$\eta_C=1-\frac{T_c}{T_h}$~. Therefore, during the optimization process, we maintain the relation~ $\frac{h_{min}}{h_{max}}=\frac{T_c}{T_h}$~, so that the optimized system can reach Carnot efficiency \cite{Ye22}. So for a given values of $T_c$ and $T_h$ we can determine the boundary values of the field.
This section examines a class of protocols involving some free parameters namely $a$, $b$, $c$, and $d$. We determine the values of these parameters by the process of optimization.  We allow the protocols to be discontinuous at $\tau/2$ and $\tau$.  Specifically, the values of the field $h(t)$ are as follows: at $t=0$, the value is $a$; at $\tau/2^{-}$, it is $b$; at $\tau/2^{+}$, it is $c$; and finally, at $\tau$, the value is $d$. 

Firstly, we take the piecewise constant protocol. This protocol is taken because as the previous result shows, this has the capability to reach the maximum achievable efficiency for the given bounds on the field magnitude. When the field is constant during each isothermal branch, the response of the system quickly stabilizes and the work extracted is maximum \cite{Ye22}. Secondly, we study the piecewise linear protocol. This protocol is a generalized version of the protocol studied in the sec. \ref{sec:Analytics_Method}. We allow for discontinuities to happen at $t=\tau/2$ and $t=\tau$.
Thirdly, we take the inverse quadratic protocol. This protocol has been observed \cite{Seifert07} to be the optimal protocol with respect to work, when the cycle time is large enough for the system to momentarily attain steady state. Apart from these, we study a fourth protocol, having the sinusoidal form. In earlier work \cite{Seifert16} the optimisation of heat engines was done by considering a sinusoidal protocol for the temperature. Motivated by this, we consider sinusoidally varying field. Another interesting aspect about the sinusoidal protocol is its non-monotonic nature, it is able to make the system release some heat for a small sub-interval even when it is in contact with the hot bath. Similarly, even when the system is in contact with the cold bath, there is a small amount of heat absorbed initially. This is not the case in other protocols considered. Instead of constraining the control, if the response of the system is constrained, optimal protocol for efficiency at fixed power takes an exponential form \cite{Dechant16}. We note that for our system, the exponential protocol gave results similar to the inverse quadratic protocol. The expressions for each protocol are provided in Table \ref{TableI}. Note that the parameters $a$, $b$, $c$, and $d$ are related to the maximum and the minimum values, the magnetic field can take
at the beginning, in the middle and at the end of an engine cycle of duration $\tau$.
\begin{table}[!t]
\begin{center} 
\begin{tabular}{ 
|p{3.2cm}|p{1.2cm}|p{2.5cm}|p{3.5cm}| p{4.5cm}|}
 \hline
 \multicolumn{5}{|c|}{List of Protocols used in the text} \\
 \hline
 Protocol duration  & PWC  & PWL    &Inverse Quadratic  &Sinusoidal \\ 
 \hline
$ 0 \leq t \leq \tau/2$  & $ a$    & $a+\frac{(b-a)t}{\tau/2} $   &$\frac{a}{\left(1+\frac{\left(\sqrt{\frac{a}{b}}-1\right)t}{\tau/2}\right)^2}$& $a+(b-a)\sin{\left(\frac{2\pi t}{\tau}\right)}$   \\
 $ \tau/2< t \leq\tau $& $ c$ & $c+\frac{(d-c)(t-\tau/2)}{\tau/2}$ & $\frac{c}{\left(1+\frac{\left(\sqrt{\frac{c}{d}}-1\right)(t-\tau/2)}{\tau/2}\right)^2}$&$c+(d-c)\sin{\left(\frac{2\pi (t-\tau/2)}{\tau}\right)}$   \\
 \hline
\end{tabular}
\caption{Functional forms of certain protocols previously examined in literature~\cite{Ye22}~in addition to the sinusoidal protocol.}
\label{TableI}
\end{center}
\end{table}
First, we discuss the piecewise constant (PWC) protocol, for which analytical expressions of heat work, and efficiency are already known and given by Eqs. \eqref{eq:PWCW}, \eqref{eq:PWCeffi}. We further extend our analysis to evaluate the work, heat, and efficiency for various other protocols, including piecewise linear (PWL), inverse quadratic, and sinusoidal protocols. The details related to actual form of these protocols are mentioned in Table \ref{TableI}. To calculate these average quantities in the long cycle time limit, we employ Eq. \eqref{eq:general_h(t)_work} and Eq. \eqref{eq:general_heat_calculation}. We discuss all the  results and compare then in the Sec. \ref{sec:Comparison_all_protocol}. But below we first discuss details about the optimization procedure used in this analysis. 
\subsection{Optimization to optimal protocol using Gradient Descent: Numerical results} \label{Optimization_using_GD}
In this part of our study, we use numerical optimization tools to attain the maximum value of the efficiency and later, power, for the spin system. We explore various protocols and observe how they approach the well-established maximum efficiency protocol through optimization. This we achieve by parameterizing the functional form of the control field, using four parameters $a$, $b$, $c$, and $d$ and optimizing these. Throughout the optimization process, we adjust the field $h(t)$ while varying free parameters to enable the system to function as an engine. Following the definition of the first law of thermodynamics (as shown in Eq. \eqref{firstlaw1}), the total extracted work from the system is taken as negative, denoted as $W < 0$. The total heat absorbed by the system is always positive, indicated as $Q_1>0$. The instantaneous heat absorption rate is always positive during the first half and always negative during the second half when the functional form of the field protocol is monotonic. When the protocol is sinusoidal, the instantaneous heat absorption rate can briefly turn negative during the first half and briefly positive during the second half. However, the net heat absorbed is still positive, and we denote that as $Q_1$. We use a simple gradient descent algorithm to optimize the efficiency and power. The procedure is described below, and a pseudo-code is provided in the Appendix \ref{appendix_GD} to describe the algorithm for work optimization. We set the temperature for both the hot ($T_h$) and cold ($T_c$) reservoirs. We also impose constraints on the field $h(t)$ to ensure it remains within a specified range. This range can be fixed at some reasonable values, which we have taken to be $0.2$ and $0.8$, without the loss of generality. We initialize the free parameters for each protocol. Then, we evaluate the work $W$ extracted from the system and the heat $Q_1$ absorbed by the system using the methods and formula described in the subsection \ref{sec:Analytics_Method}. Following this, we calculate the extracted power and the corresponding efficiency. Next, we fix the quantity we want to optimize for our study: efficiency or power. Then we increment and decrement each parameter, and sample the corresponding values of the quantity. Then we find the corresponding component of the gradient vector $\vec{g}$ from the two data points with the central difference formula. We then, update the values of all the parameters in the direction of the gradient vector scaled by a ``learning rate parameter" $\alpha$ chosen suitably to achieve the desired accuracy and the computational expense incurred. 
\begin{equation}\label{updation_equation}
    [a,b,c,d] \rightarrow [a,b,c,d] + \alpha \vec{g} .
\end{equation}
Then, the process of incrementing and decrementing the parameters is repeated. When the norm of the gradient vector is smaller than a threshold value, we stop the optimization process and designate the protocol and efficiency or power as optimal.
\subsubsection{Optimizing for Efficiency}
\label{subsection:optimizing_efficiency}
\begin{figure}[!thbp]
\includegraphics[width=0.45\textwidth]{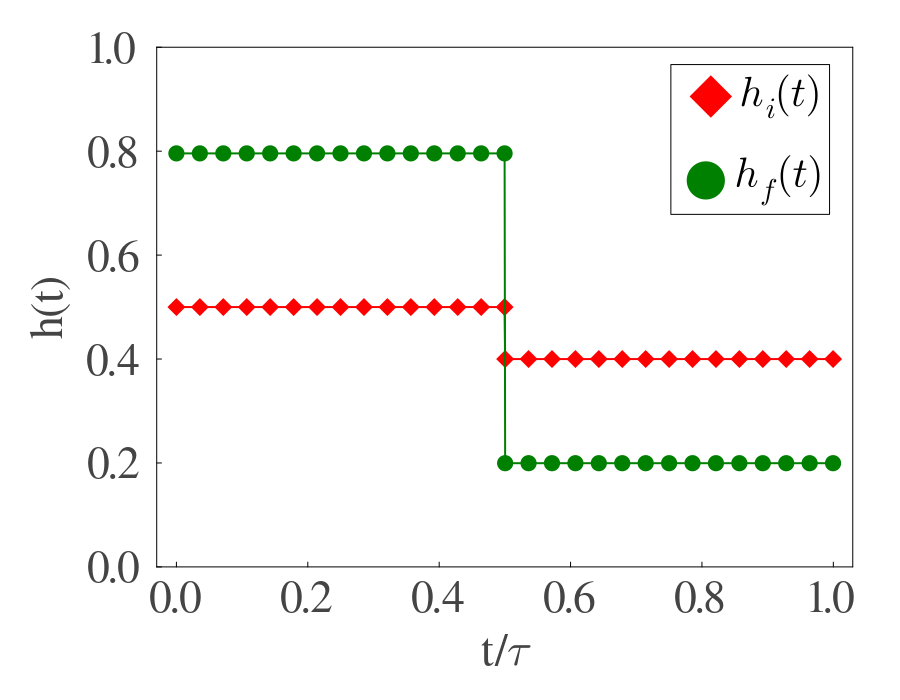}   \includegraphics[width=0.45\textwidth]{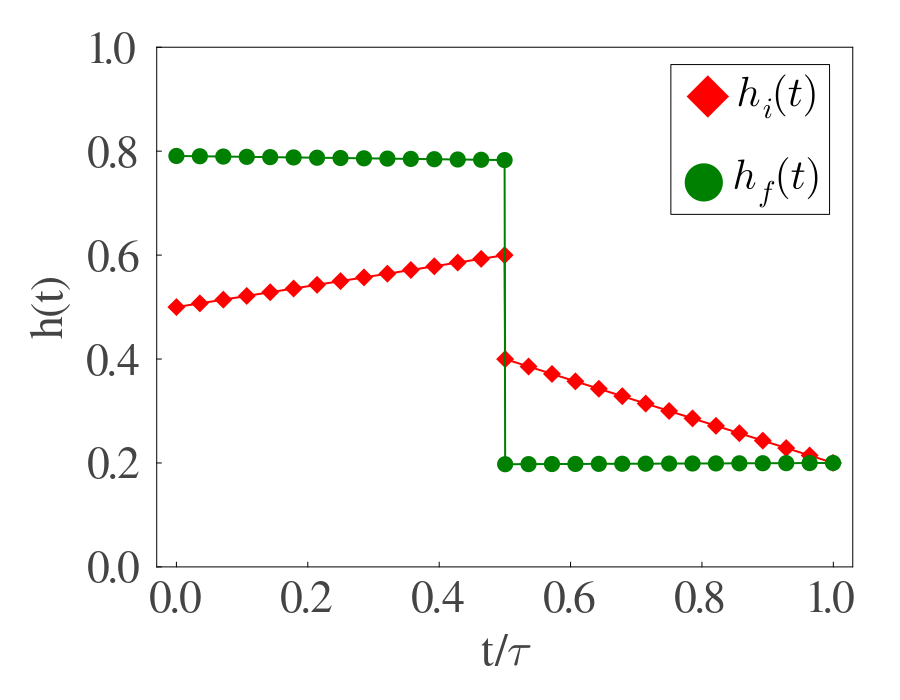}
    \includegraphics[width=0.45\linewidth]{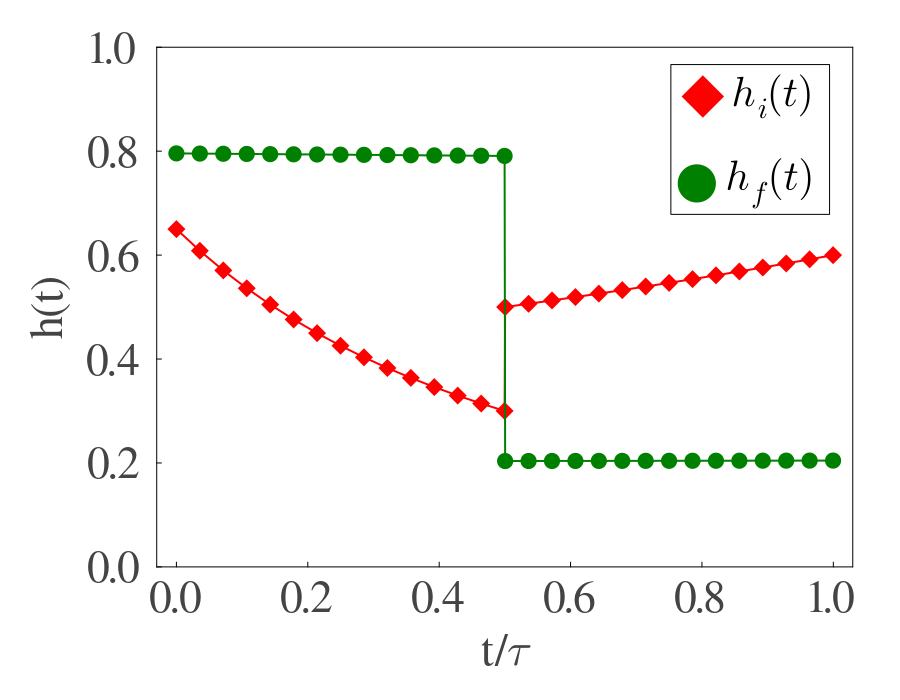}
    \includegraphics[width=0.45\linewidth]{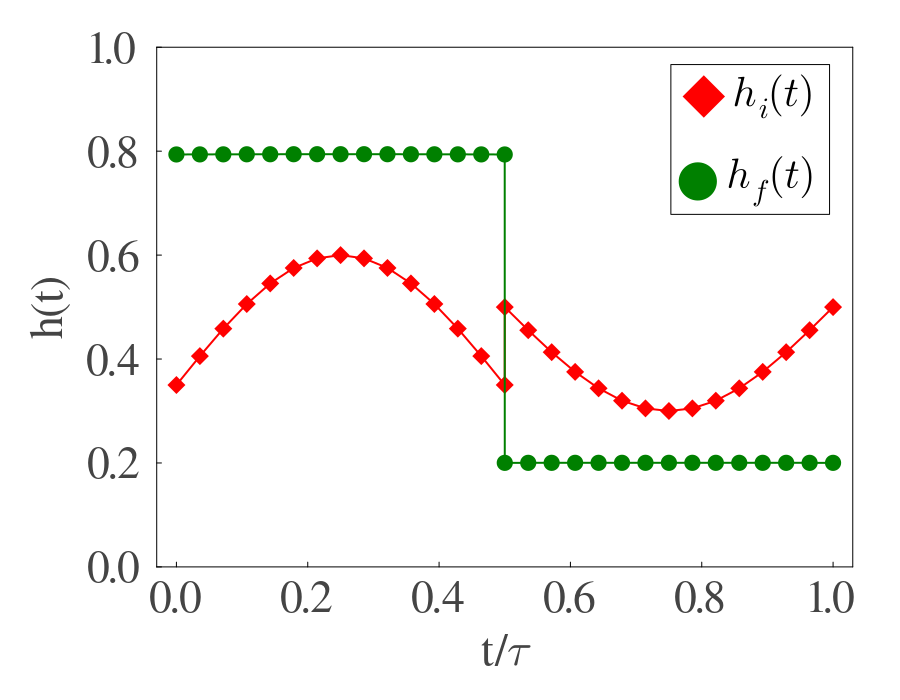}
    \caption{Numerical optimization of the efficiency by optimizing the free parameters $a$, $b$, $c$, $d$. Within each plot displayed in the four images, we represent the initial protocol  in red ($h_{i}(t)$). At the same time, the green color corresponds to the final protocol type ($h_{f}(t)$) after the efficiency optimization. The plots represent the, (top left): Piecewise Constant (PWC),  (top right): Piecewise Linear (PWL), (bottom left): Inverse quadratic and (bottom right): Sinusoidal. The temperatures are set $T_c=0.5, T_h=2$, the maximum value of the field $h(t)$ is $h_{max}=0.8$ and $h_{min}=0.2$ sets as a constraint. The initial values we choose are  for PWC ($a=0.5$, $c=0.4$), PWL ($a=0.5$, $b=0.6$, $c=0.4$, $d=0.2$), Inverse quadratic ($a=0.65$, $b=0.3$, $c=0.5$, $d=0.6$), and sinusoidal ($a=0.35$, $b=0.6$, $c=0.5$, $d=0.3$). The final value after optimization is the same for all because all reached the maximum efficiency protocol with optimal parameter $a=b=0.8$, $c=d=0.2$}
     \label{fig:optimize_effi}
\end{figure}
\begin{figure}[!thbp]
\includegraphics[width=8.5cm,height=6.1cm,angle=0]{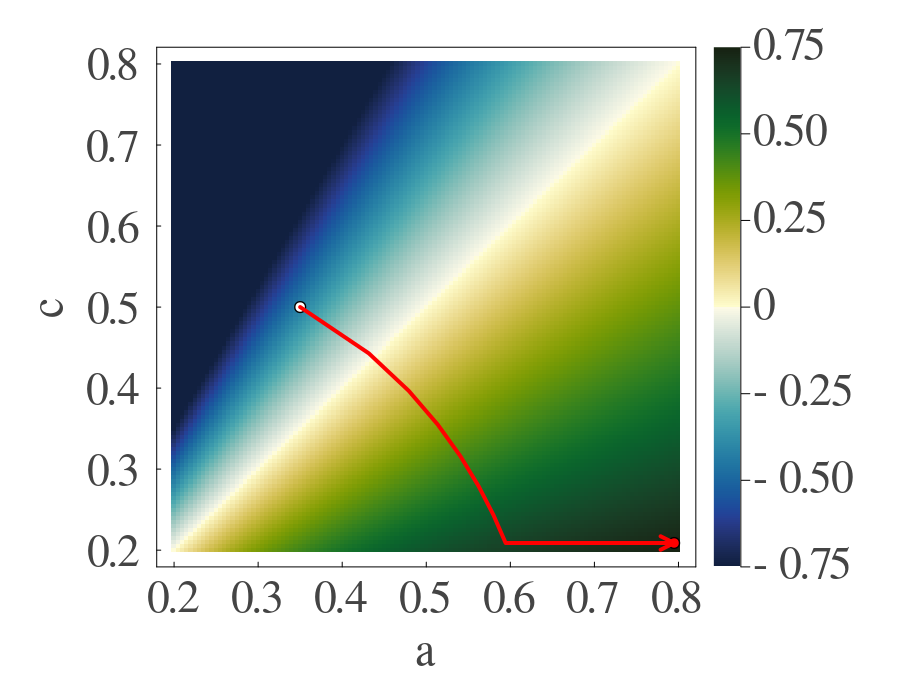}
\quad
\includegraphics[width=8.1cm,height=6.1cm,angle=0]{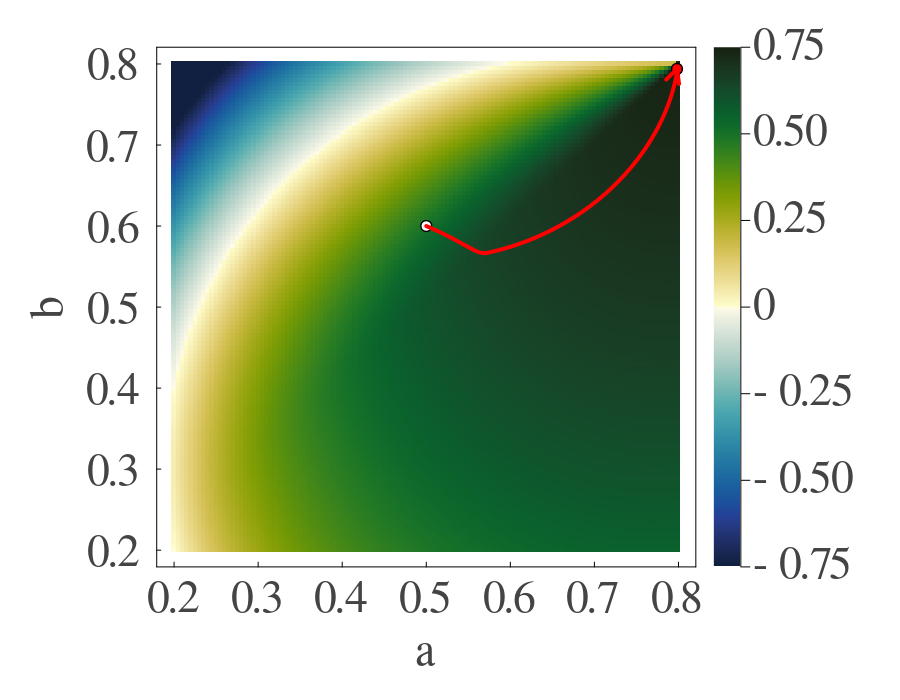}
\includegraphics[width=8.1cm,height=6.1cm,angle=0]{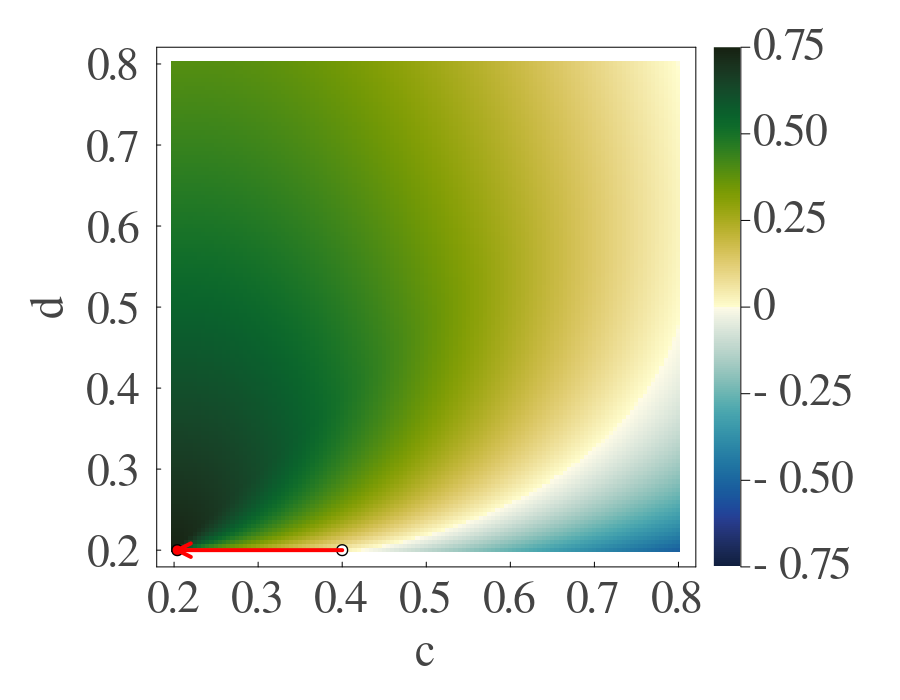}
\caption{(top) Phase plot for efficiency with free parameters $(a,c)$ for the PWC protocol. the initial parameters are $a=0.5$, $c=0.4$. (bottom left) and (bottom right) Pair plots for the PWL protocol. The plots show the different efficiency values with variation with parameters ($a,b$) while $ c=d=0.2$ is fixed at optimal value and the same with ($c,d$) while $a=b=0.8$ is fixed at optimal value. For each  plot above, Other parameters are $T_c=1$, $T_h=4$, $\mu=1$. We denote the parameter's initial values or starting point of the trajectory by the white dot. The optimal protocol values after the optimization process are defined as the red arrow in the parameter space.} 
\label{fig:phaseplot:optimize_efficiency}
\end{figure}
In this part, we  discuss the results we obtain from the optimization process. Fig. \ref{fig:optimize_effi}, shows the evolution of protocols ultimately leading to the optimized protocol. In each of these plots, we represent the initial protocol type in red ($h_{i}(t)$), while the green color corresponds to the final protocol type ($h_{f}(t)$). The PWC protocol (top left) given in the first column of table \ref{TableI}, is the simplest one and is also the maximum efficiency protocol. In this case, within the interval $0 < t < \tau/2$, the field $h(t)$ maintains a constant value of $a$, while within the interval $\tau/2 < t < \tau$, it assumes the value $c$. We start with the initial values of the parameters of the protocol  $a=0.5$, $c=0.4$, chosen arbitrarily. After optimization, it closely approaches the constraining boundary values of $0.8$ and $0.2$. In the PWL protocol, given in the second column of table \ref{TableI}, we vary the field linearly with time. We initialize the protocol with the parameters $a=0.6$, $b=0.6$, $c=0.4$, $d=0.2$.  These parameters represent the initial and final values of the field in each branch of the cycle. So the field goes from $a$ to $b$ linearly, jumps to $c$ and then goes to $d$ linearly in Fig. \ref{fig:optimize_effi} (top right). After optimization, the protocol reduces to a PWC protocol which reaches the boundary values. This verifies our earlier result, that this PWC is the most efficient protocol. This is apparent from the optimized values of the system parameters: $a=b=0.8$, $c=d=0.2$. The process is exactly same for the inverse quadratic protocol, given in the third column of table \ref{TableI}, which again reduces to the PWC form in Fig. \ref{fig:optimize_effi} (bottom left). The optimization procedure for the sinusoidal protocol, given in the fourth column of table \ref{TableI} slightly differs from others. Since the sine function is non-monotonic nature, its extreme values do not occur at the endpoints of the respective half-cycles.  The parameter $a$ and $c$ still represent the initial values of the field for each branch, but the parameters $b$ and $d$ represent the value of the field at the midpoint of each half-cycle. This new definition of the parameters as well as due to the overall shape of the protocol has a peculiar consequence: its convergence  to the PWC protocol in Fig. \ref{fig:optimize_effi} (bottom right) requires a smaller step size and more computational power in the simulation compared to the other protocols. 

In Fig. \ref{fig:phaseplot:optimize_efficiency} (top), we present a phase plot illustrating the efficiency of the Piecewise Constant (PWC) system, with the control parameters denoted as $a$ and $c$. We aim to observe how the efficiency evolves as we manipulate these two parameters, striving to reach the optimal value. (bottom left) and (bottom right) shows two pair plots for PWL protocol. Given that the parameter space is four-dimensional, our visualization approach involves varying only two parameters at a time within these plots while keeping the other two sets at their optimal values. It should be noted that our algorithm optimizes all parameters simultaneously. This method demonstrates the system's progression towards achieving Carnot efficiency. The area shaded in yellow-green represents the region where the system functions as an engine, while the bluish region indicates when it does not. The parameters transit from their initial values towards reaching the optimal configuration with $a=0.8$ and $ c=0.2$.
\subsubsection{Optimizing the Power output}
\label{subsection:optimizing_power}
In this part, we maximize the power output. We calculate the power using Eq. \eqref{eq:power_entropy_all}, but finding the analytical expression of optimal protocols is a very challenging task. But numerically, we calculate the optimized power by the gradient descent algorithm. We observe that when the power is optimized by this method, the value of the efficiency is always bounded above by the Curzon-Ahlborn efficiency limit ($\eta_{CA} = 1-\sqrt{\frac{T_c}{T_h}}$) \cite{Curzon}. This, for the values of boundaries and temperatures chosen, is $1-\sqrt{\frac{1}{4}}=0.5$.
\begin{figure}[!thbp]
     \centering     \includegraphics[width=0.45\linewidth]{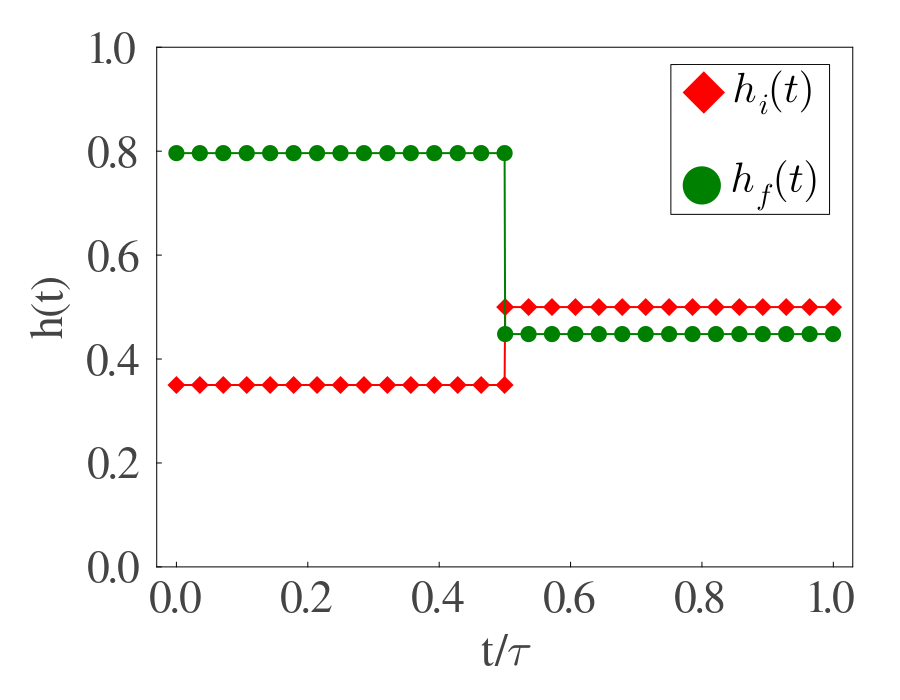}
     \includegraphics[width=0.45\linewidth]{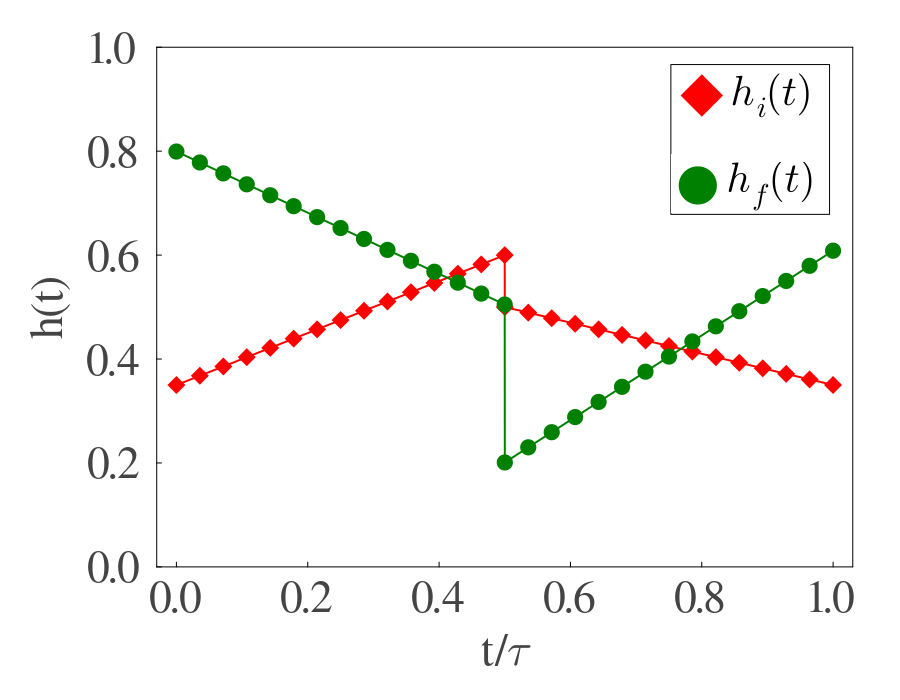}
     \includegraphics[width=0.45\linewidth]{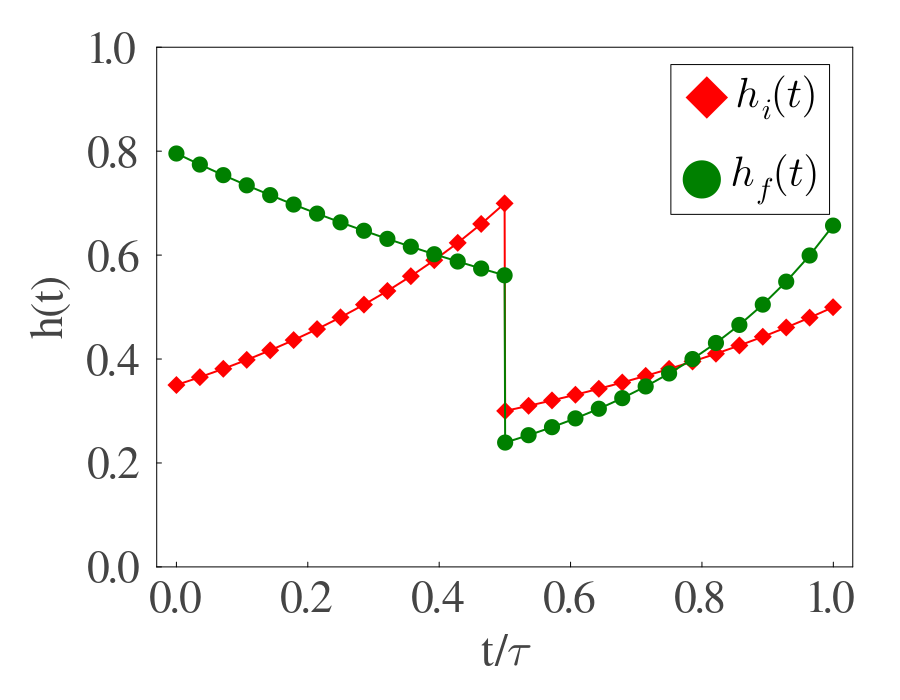}
     \includegraphics[width=0.45\linewidth]{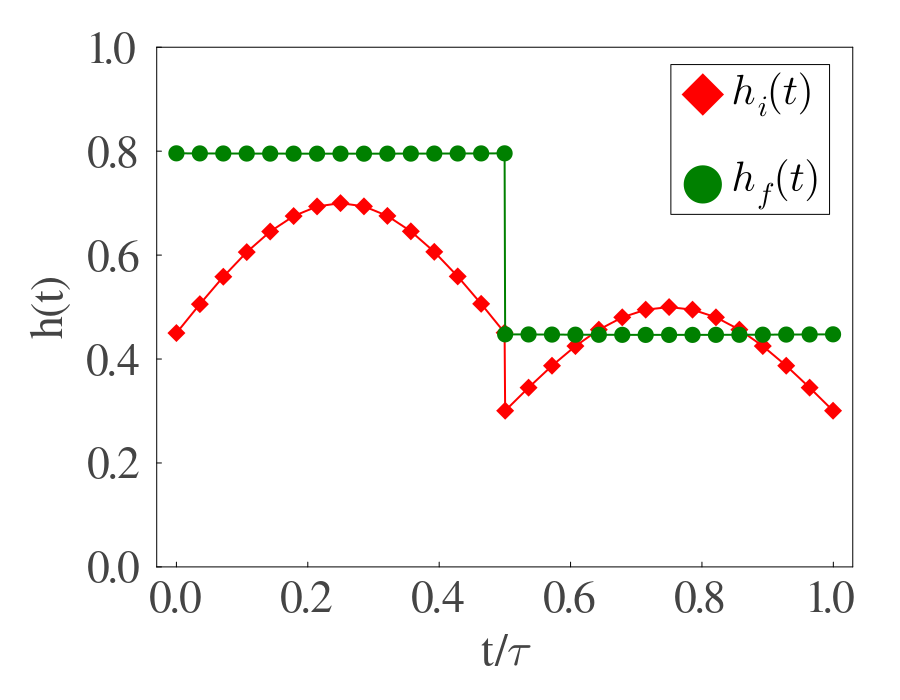}
     \caption{ Numerical optimization of the output  power by optimizing the protocols free parameter $a$, $b$, $c$, $d$. Within each plot displayed in the four images, the initial protocol type is represented in red ($h_{i}(t)$). At the same time, the green color corresponds to the final protocol type ($h_{f}(t)$) after the efficiency optimization. The plots represent the,
     (top left) Piecewise Constant (PWC),  (top right) Piecewise Linear (PWL), (bottom left) Inverse quadratic and (bottom right) Sinusoidal. The temperatures are set $T_c=0.5, T_h=2$. The initial values we choose are  for PWC ($a=0.35$, $c=0.4$), PWL ($a=0.35$, $b=0.6$, $c=0.5$, $d=0.35$), Inverse quadratic ($a=0.35$, $b=0.7$, $c=0.3$, $d=0.5$), and sinusoidal ($a=0.45$, $b=0.7$, $c=0.3$, $d=0.5$).}
     \label{fig:power_protocol}
      \end{figure}
In the Fig. \ref{fig:power_protocol}, we optimize the output power by fine-tuning the  parameters $a$, $b$, $c$, and $d$. In each of these plots, we again represent the initial protocol type in red ($h_{i}(t)$), while the green color corresponds to the final protocol type ($h_{f}(t)$). Our optimization process begins with an initial set of free parameters, and through numerical methods, it adjusts these parameters to optimize the power output. As a result, we ultimately determine the final protocol type. The figure shows that the PWC protocol (Fig. \ref{fig:power_protocol} (top left)) does not reach the boundary values ($a=0.8, c=0.2$). The first branch maximizes the input heat. However, the subsequent branch adopts a trajectory that leads the system towards the values that maximize the power. When considering the PWL protocol (Fig. \ref{fig:power_protocol} (top right)), the initial configuration demonstrates that the function behaves as an increasing function during the first half of the cycle. However, following an adiabatic jump, it transforms into a decreasing function. This behavior doesn't align with the engine behavior we typically expect. Upon optimizing for power, the protocol undergoes a significant transformation. Its nature changes to become a decreasing function during the first half and subsequently transforms into an increasing function in the second half of the cycle. This alteration in behavior not only signifies engine-like characteristics but also reflects the achievement of maximum power output. The inverse quadratic protocol's behavior is similar to the piecewise linear protocol (Fig. \ref{fig:power_protocol} (bottom left)). However, the efficiency value at maximum power is better than the PWL. The sinusoidal protocol (Fig. \ref{fig:power_protocol} (bottom right))converges to the Piecewise Constant (PWC) protocol when we optimize the power. The shape of the maximum power protocol requires different endpoints of each branch. While the sinusoidal and PWC has same endpoint for each branch unlike PWL and inverse quadratic protocol. So these protocols show discontinuity and attain the desired parameter value which makes the efficiency approach the Curzon-Ahlborn limit.
\begin{figure}[!thbp]
\centering
\includegraphics[width=8.5cm,height=6.1cm,angle=0]{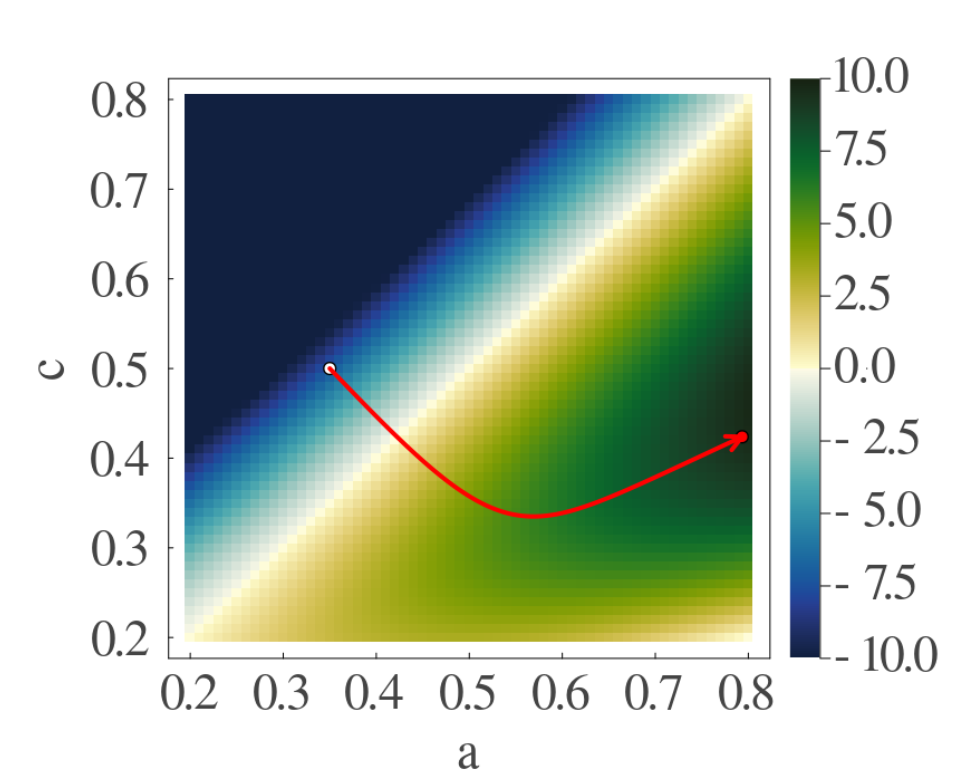}
\quad
\includegraphics[width=8.1cm,height=6.1cm,angle=0]{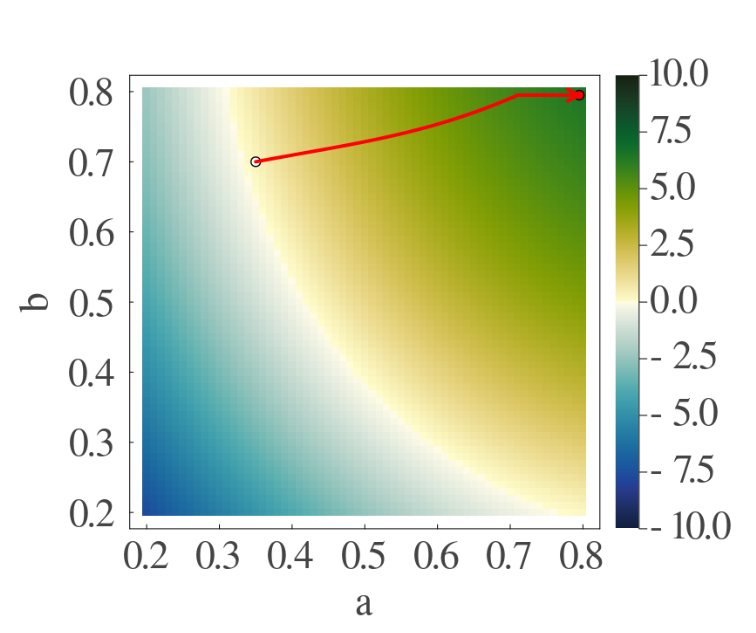}
\includegraphics[width=8.1cm,height=6.1cm,angle=0]{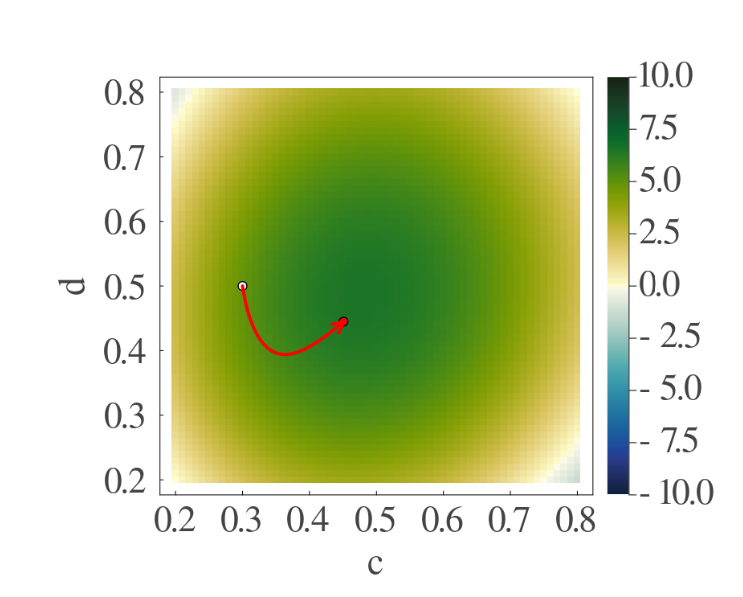}
\caption{ (top) Phase plot for power (to be multiplied by $10^{-3}$) with free parameters $(a,c)$ for the PWC protocol. The initial parameters are $a=0.35$, $c=0.5$. (bottom left) and (bottom right) Pair plots for the sinusoidal protocol. The plots show the different power values with variation with parameters ($a,b$) while $ c,d$ is fixed at optimal value and the same with ($c,d$) while $a,b$ is fixed at optimal value. For each  plot above, Other parameters are $T_c=1$, $T_h=4$, $\mu=1$. We denote the parameter's initial values or starting point of the trajectory by the white dot. The optimal protocol values after the optimization process are defined as the red arrow in the parameter space.}
\label{fig:phaseplot:optimal_poweroutput}
\end{figure}
 In Fig. \ref{fig:phaseplot:optimal_poweroutput} (top), we provide a phase plot depicting the output power of the Piecewise Constant (PWC) protocol, where the free parameters are denoted as $a$ and $c$. We aim to observe how the output power evolves as we manipulate these two parameters to reach an optimal value. This effectively illustrates the system's progression towards achieving maximum power. The shaded yellow-green region signifies where the system operates as an engine, while the bluish region indicates when it does not. This is followed by the phase plots depicting the power output of the sinusoidal protocol. Firstly, the optimization is done numerically and the values of the optimal parameters are noted down. The Fig. \ref{fig:phaseplot:optimal_poweroutput} (bottom left) shows the variation of power as a function of (a,b) when (c,d) are held fixed at their optimal values. Then the Fig. \ref{fig:phaseplot:optimal_poweroutput} (bottom right) shows the variation of power vs (c,d) when (a,b) are held fixed at their optimal values. These plots show an interesting circular symmetry. 
 This optimization method requires a functional form of the protocol that can be fitted. We now ask, can a more general form of the protocol be derived (without assuming any initial form) for efficiency and power optimization? 
 In the next Sec.  \ref{variational}, we explore this question by employing a general methodology to optimize the efficiency and power using a variational iterative method.

\section{General Optimal protocols for Power and efficiency using Variational iterative method} \label{variational}
Until now, we have used an unconventional method to find optimal protocols for the field by fixing the form of functions we believe can represent the optimal function and varying their endpoints (or mid-points for sinusoidal functions) to achieve the best fit. However, there is a more general and accurate method for finding the optimal protocol, detailed in \cite{CraunOptimalJDyn15, Symmetry}, which optimizes the power and efficiency using a variational method. The procedure to do this is as follows. First, we define the quantity to be optimized as an integral over the cycle, with the integrand denoted by \(\xi\). The system's dynamical equation is written as \(\dot{m} = f\), where \(m\) represents the system's response to the control field \(h(t)\). Define the Hamiltonian \(H(m,h,\lambda) = \xi(m,h) + \lambda(t) f(m,h)\). Here, $\lambda$ serves as a Lagrange multiplier. Since the constraint from the dynamical equation is to be followed at every point of time, the multiplier becomes a function of time. Then, we derive the equation of motion for \(\lambda\) as \(\dot{\lambda} = -\frac{\partial H}{\partial m}\) and determine the increment of the control field as \(\delta u = \frac{\partial H}{\partial h}\).  In each iteration, we solve for \(\lambda(t)\), increment the control field, and use the updated control field to solve for the new system response \(m\) through the dynamical equation. This iterative process allows for a much more accurate optimization of the system's power and efficiency.
\subsection{Power Optimization}\label{variational_power}
\subsubsection{Hard Boundary Conditions}
To optimize power, we focus solely on optimizing the extracted work, given that the cycle time is fixed. The dynamical equation for our system is,
\begin{align}
    \dot m(t) = -m(t)+ \tanh(\beta(t) h(t)) \label{eq_dyna} 
    \end{align}
    Where we have taken $\mu=1$. Here, $\beta(t) = \frac{1}{T(t)}$, with the temperature protocol being piecewise constant in each half cycle, $T_h$ for $0 \leq t \leq \tau/2$, and $T_c$ for $\tau/2 \leq t \leq \tau$. 
The rate of work is given by \(\dot{W} = -\dot{h} m\), and the rate of heat absorption is \(\dot{Q} = -h \dot{m}\). To maximize the total work extracted, we rewrite the expression of extracted work as \(W_{ex} = -\int_0^\tau \dot{W} \, dt\). Thus, we write the target integrand as,
\begin{align}
        \xi(m, h)= h(m-\tanh(\beta h))
    \end{align}
Then we add the constraint term to it and write the evolution equation for $\lambda$ as, 
\begin{align}
\dot \lambda = -h + \lambda \label{eq_lambda_pow}
\end{align}
We write the increment to be made to the field at each iteration is by,
\begin{align}
\delta h = m - \tanh(\beta h) + (\lambda - h) \beta \text{ sech}^2(\beta h) \label{eq_h_pow}
\end{align}
\begin{figure}[!t]
\centering
\includegraphics[width=0.5\linewidth]{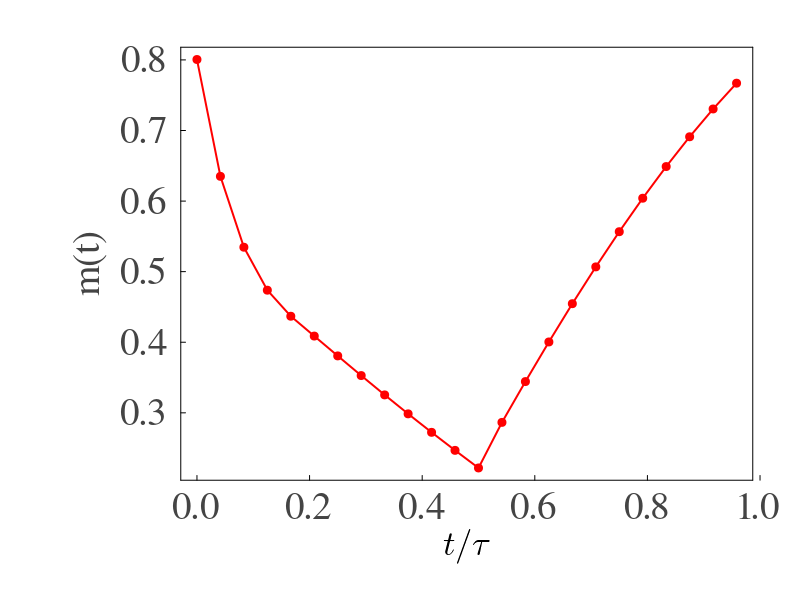}\hfill
\includegraphics[width=0.5\linewidth]{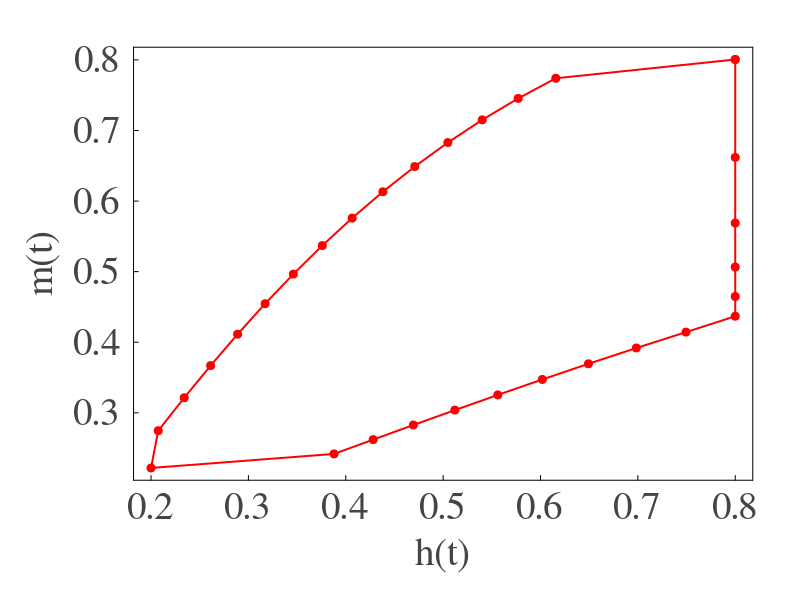}\hfill
\includegraphics[width=0.5\textwidth]{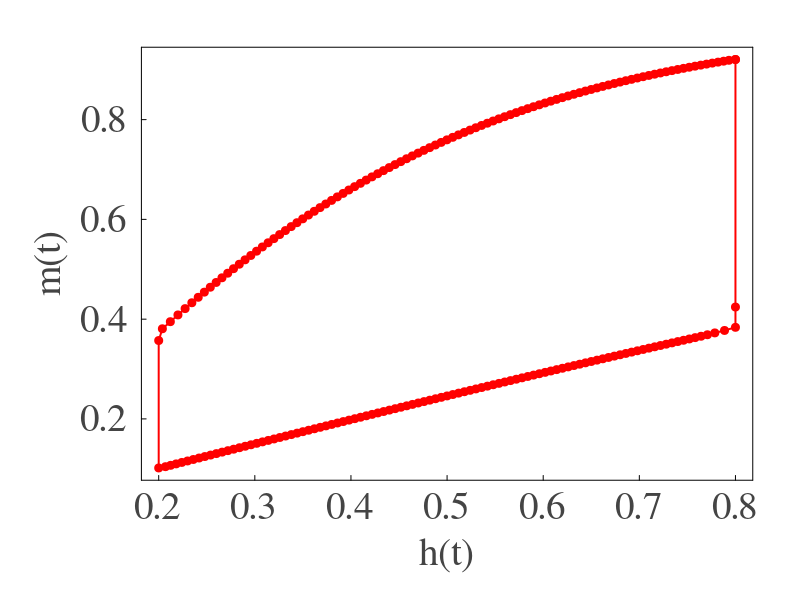}\hfill
\includegraphics[width=0.5\linewidth]{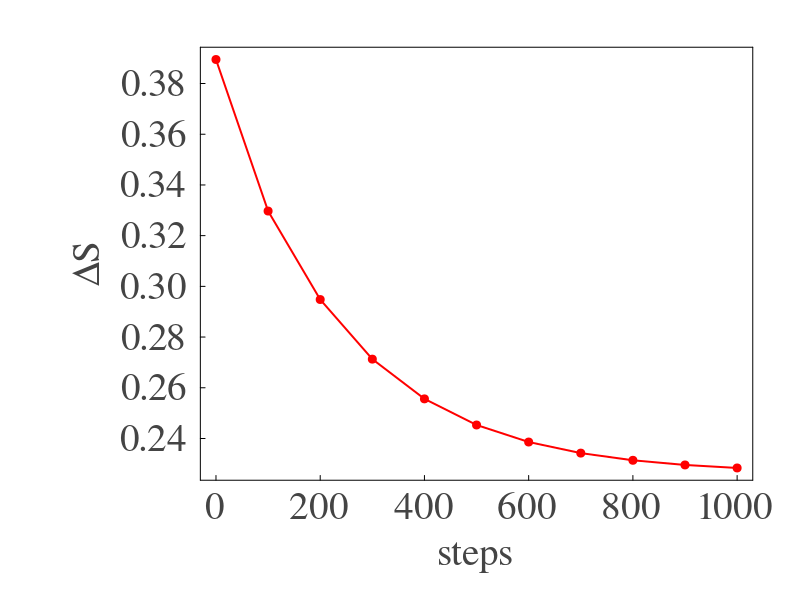}
\caption{
(top left) Represents magnetisation $m(t)$ with time $t$ for optimal power protocol. 
(top right) Shows magnetization $m(t)$ with field $h(t)$ for optimal power protocol for $\tau=12$. (bottom left) Magnetization $m(t)$ with field $h(t)$ for a larger cycle time limit, $\tau=500$. (bottom right) Total entropy production $\Delta S$ with number of iterations.}
\label{mag_vs_h}
\end{figure}
We provide a pseudo-code in Appendix \ref{Varalgo} to describe the algorithm used for optimizing work.
For the optimal power protocol, in Fig. \ref{mag_vs_h} (top left), we show the variation of the magnetization \(m(t)\) with time \(t\). Immediately after the temperature switches to the hot bath, there is a rapid increase in temperature-induced fluctuations, causing the magnetization to decrease. A reduction in the magnitude of the field follows the constant portion, but this does not lead to any noticeable further change in the magnetization. This indicates that the effect of the temperature shift is more significant than that of the field shift in the initial portion of the cycle. Subsequently, as we decrease the field, the magnetization also decreases.
In the second half of the cycle, an increase in magnetization due to a decrease in temperature-induced fluctuations is observed. After a small constant portion, the field increases, though not completely linearly, and thus, the increase in magnetization is not linear either. To properly illustrate the control-response dependence, we plot the magnetization versus the field $h(t)$ in the (top right) and compare the plot of $m(t)$ with the field $h(t)$ for the large cycle time. It is evident from the plot in Fig. \ref{mag_vs_h}~(bottom left) that the cycle of the magnetization versus field tends to attain the shape of the Stirling cycle as we increase the cycle time.

We now analyze the total entropy production $\Delta S$, during the optimization process. We define the thermodynamic entropy production as,
\begin{equation}
\Delta S = -\int_0^{\tau} \frac{\dot Q(t)}{T(t)} ~dt.
\end{equation}
For the optimized power protocol, there is always a finite amount of entropy production for a given cycle time, as shown in Fig. \ref{mag_vs_h}~(bottom right). The magnetization remains unchanged when the ratio of the fields matches the ratio of the temperatures before and after the jump (i.e. when $\beta_h$ is constant). However, this is not the case for the maximum power protocol, which significantly changes the system's magnetization. This leads to the generation of waste heat, a reduction in efficiency, and an increase in entropy production. The total optimized entropy production for our system is $0.228$ units. To achieve this, we begin with a piecewise linear trial protocol and observe the changes in entropy production during the optimization process. The protocol with hard boundary conditions (for a cycle time of $\tau=12$) yields better power optimization compared to the results obtained using the piecewise linear protocol (see Table \ref{Table_II}). Below, we analyze how allowing the bounds on the external parameters to be flexible changes the behavior of the protocol. These are termed as soft boundary conditions \cite{CraunOptimalJDyn15}.

\subsubsection{Comparison with soft boundary conditions} \label{softhard}
\begin{figure}[t]
    \centering
\includegraphics[width=0.5\linewidth]{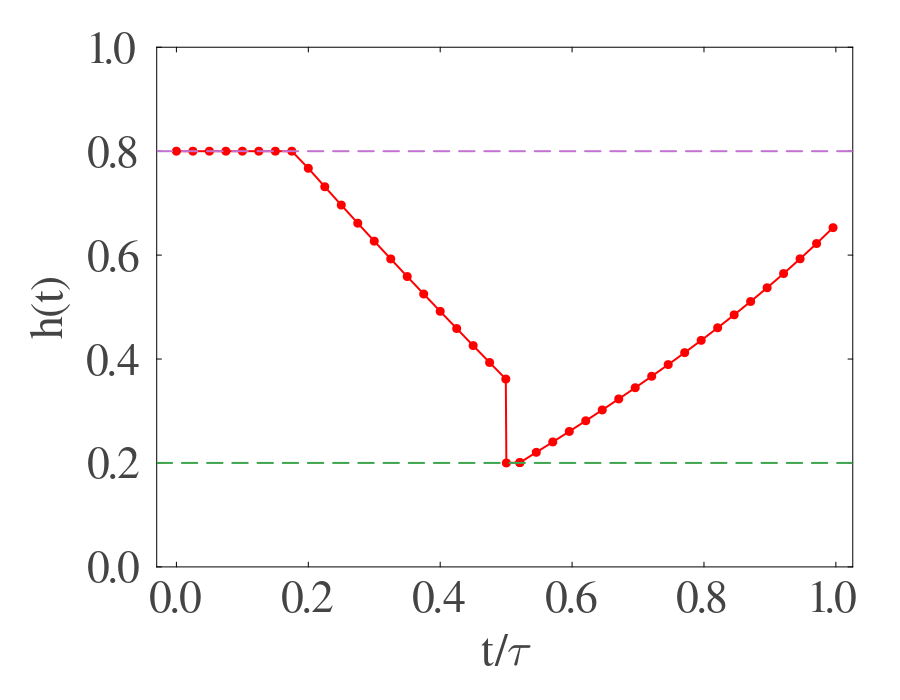}\hfill
\includegraphics[width=0.5\linewidth]{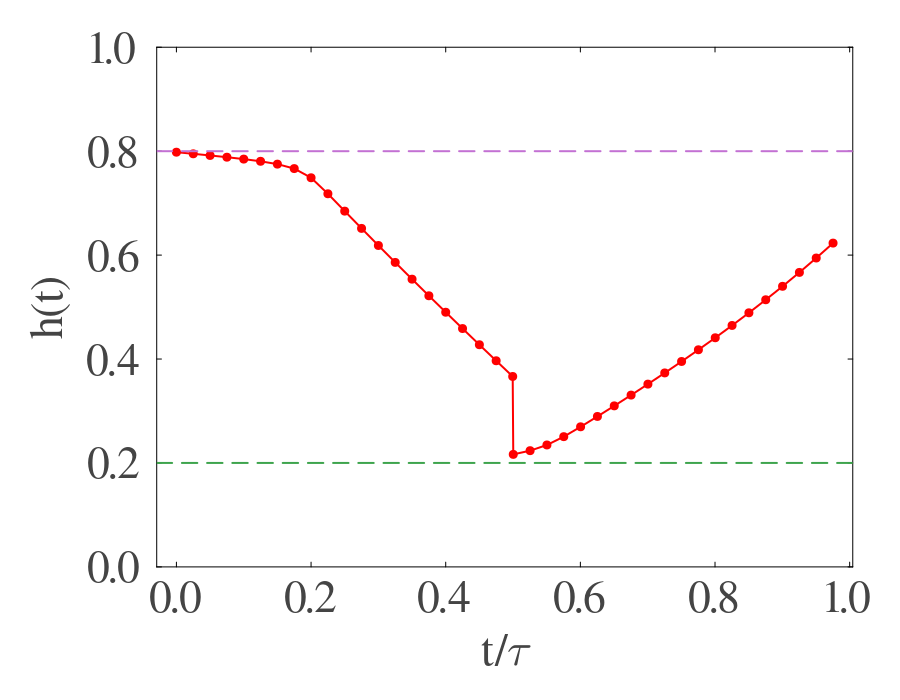}
\caption{(left) The optimized protocol for maximum power with hard constraints on the field boundaries. (right) The optimized power protocol using soft boundary conditions. The parameters used are same as in Table \ref{Table_II}, they are $\tau=12, T_{h}=2,T_{c}=0.5,h_{max}=0.8, h_{min}=0.2$.}
\label{hard_soft}
\end{figure}

The primary objective of this optimization study is to identify the optimal protocol for the magnetic field that maximizes power or efficiency for a given temperature protocol while adhering to constraints imposed on the control protocol. These constraints are applied externally during the incremental adjustment of the field $h(t)$ by $\delta h(t)$, resulting in a smooth and practical field protocol. However, this approach prevents the increment function $\delta h(t)$ from fully converging to zero. Soft boundary conditions can be introduced to achieve more natural convergence. By adding a pair of cost terms (one for each boundary) to the target function $W_{ex}$, each term remains approximately zero when the protocol is distant from the corresponding boundary but becomes significantly negative as the protocol approaches the boundary. This mechanism encourages the updating function $\delta h(t)$ to detect and avoid the boundary. The soft boundary condition can thus be expressed as
\begin{equation}
H_{soft}(m,h,\lambda)  = \xi(m,h) + \lambda(t)f(m,h) - C_0 ( e^{C_1(h(t)-0.8)}+e^{C_1(0.2-h(t))})
\end{equation}
where we adjust the value of $C_0$ and $C_1$ accordingly. Since this term only depends on $h(t)$ by modifying the $\delta h(t)$ equation as,
\begin{align}
\delta h_{soft} = m - \tanh(\beta h) + (\lambda - h) \beta \text{ sech}^2(\beta h) -C_0 C_1(e^{C_1(h(t)-0.8)}-e^{C_1(0.2-h(t))})
\end{align}
Here, $C_0=0.005$ and $C_1=100$ yield acceptable results. The optimal power protocol under both hard and soft boundary conditions is illustrated and compared in Fig. \ref{hard_soft}. This protocol converges when the equation $\delta h_{soft}(t)=0$ is satisfied. Using the soft constraint, the final form of the protocol~(Eq. \eqref{eq_h_pow}), which produces the highest power extraction, is given by Fig. \ref{hard_soft} (right) and compared with the protocol obtained using the hard constraint Fig. \ref{hard_soft}~(left). 

As expected, the power extracted using the protocol with soft constraints is lower than that obtained with hard constraints. The efficiency is approximately $0.47$, close to the Curzon-Ahlborn efficiency of $0.5$. This efficiency converges more closely to 0.5 as the cycle time increases, thereby optimizing the power output for that specific cycle time. In this study, we have fixed the cycle time at $\tau=12$, resulting in a characteristic shape of the protocol. However, it is important to note that the shape of the protocol is highly sensitive to both the boundary constraints imposed on the field and the cycle time. These constraints compress the field within the allowed region, leading to the constant segments observed in the protocol. The influence of cycle time on the protocol shape and the rigidity of the boundary conditions is analyzed in Appendix \ref{cycle_time}. As illustrated in Fig. \ref{nineplots}, increasing the cycle time reduces the magnitude of the jumps while varying the softness of the boundary conditions can induce overshoots.

\subsection{Efficiency Optimization}
The procedure for optimizing efficiency is slightly different than the power optimization process. Since efficiency is the ratio of two functionals, we need to consider the variation in efficiency carefully. First, we take the partial derivative of both sides of the efficiency equation, as shown below,
\begin{align}
   \delta \eta \equiv \delta(\frac{W}{Q_+}) = \frac{\delta W}{Q_+} - \frac{W}{Q_+^2} \delta Q_+
\end{align}
where, $W$ is the total work extracted, and $Q_+$ is the total heat absorbed in a cycle. 
\begin{figure}[!t]
    \centering    
    \includegraphics[width=0.45\linewidth]{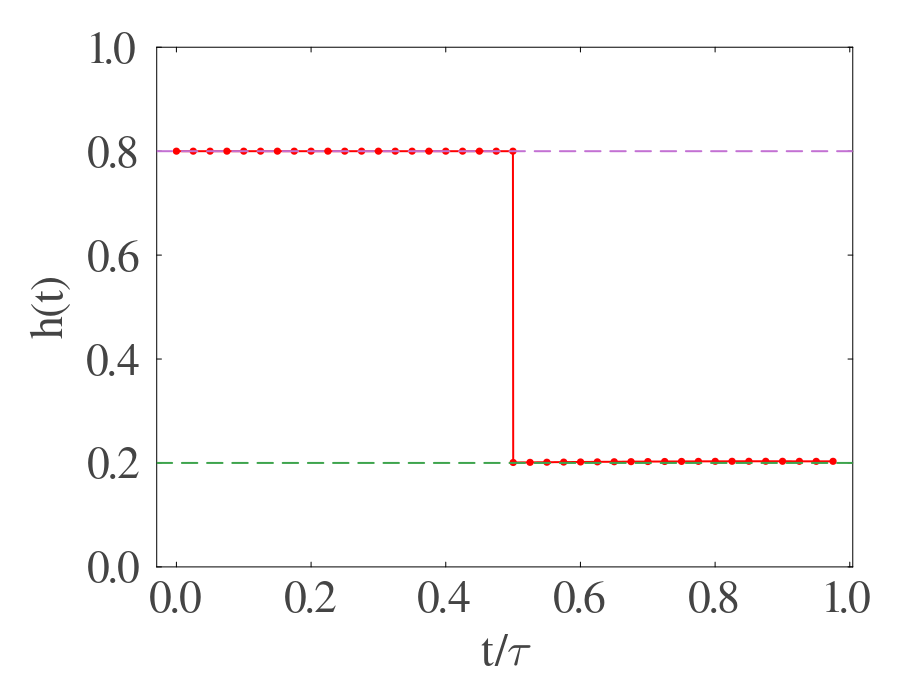}
    \includegraphics[width=0.47\linewidth]{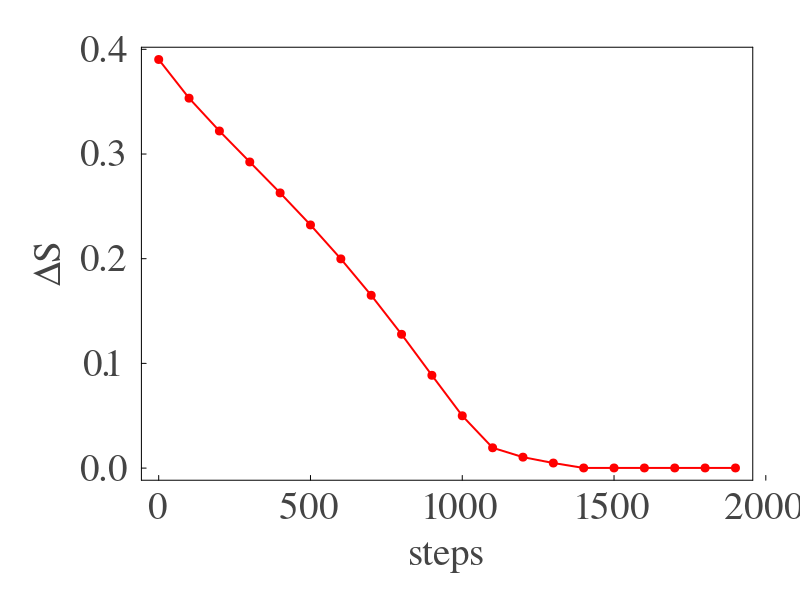}
    \includegraphics[width=0.49\linewidth]{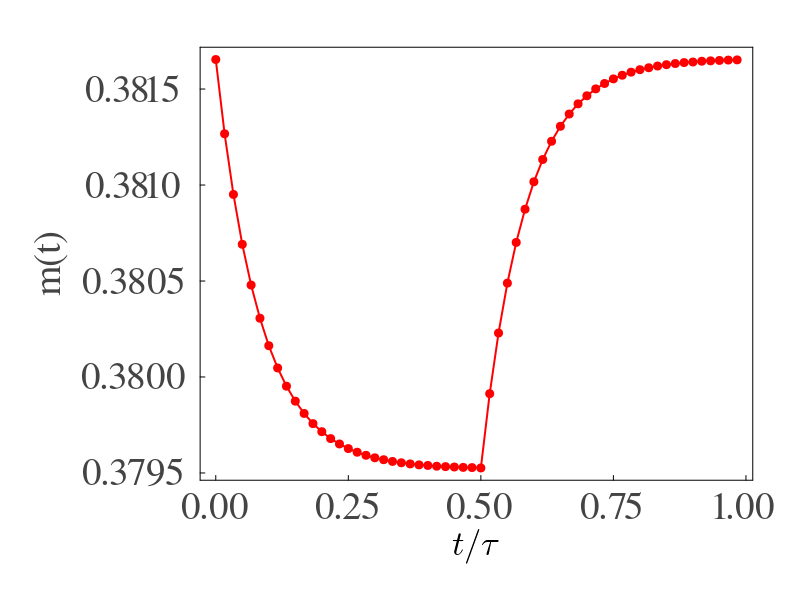}\hfill
    \includegraphics[width=0.5\textwidth]{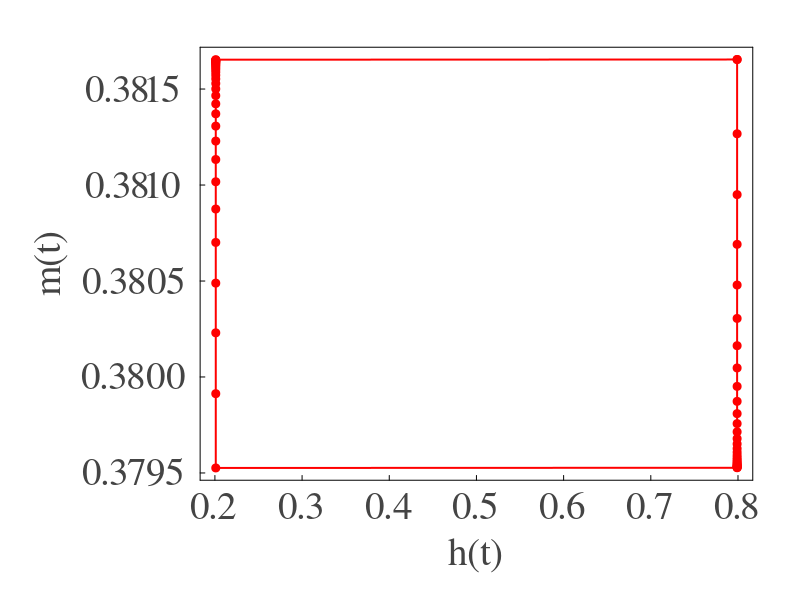}
     \caption{(top left) Optimal efficiency protocol for the same parameters as used in the Fig. \ref{hard_soft}. (top right) Entropy production with each iteration step. (bottom left) magnetization vs. time for the optimal efficiency protocol. (bottom right) Magnetization vs field plot for optimal efficiency protocol}
    \label{var_eff_proto}
\end{figure}
So, the target function is the time integral of the following,
\begin{align}
    \xi_\eta = \frac{W}{\tau Q_+} + \frac{\dot W}{Q_+} - \frac{W}{Q_+^2} \dot Q_+
\end{align}
The first term is added to prevent the second and third terms from canceling each other out after integration, which would result in zero efficiency. The inclusion of this term does not affect the optimization process since it is a constant from the previous iteration. Using the definitions of work and heat rates, we can write,
\begin{align}
\xi_\eta=\frac{W}{\tau Q_+} -\frac{h}{Q_+} (-m + \tanh(\beta h)) + \frac{h W}{Q_+^2}(-m + \tanh(\beta h))~ \theta(m-\tanh(\beta h))
\end{align}
\begin{align}
    \dot \lambda = -\frac{h}{Q_+} + \frac{hW}{Q_+^2}~\theta(m-\tanh(\beta h)) - \frac{W h}{Q_+^2}(-m + \tanh(\beta h))~\delta(m-\tanh(\beta h))+ \lambda \label{eq_lambda_eff}
\end{align}
We increment the protocol by the following,
\begin{eqnarray}
        \delta h &\propto& \frac{-\dot m -\beta h \text{ sech}^2(\beta h)}{Q_+} + \frac{W}{Q_+^2}\left (\theta(-\dot m)~( \dot m + \beta h~ \text{sech}^2(\beta h))-\beta h~ sech^2(\beta h)~\dot m~\delta(-\dot m) \right ) \nonumber\\
    &+& \beta \lambda \text { sech}^2(\beta h) \label{eq_h_eff}
\end{eqnarray}
where we have used $\dot m$ in some terms for brevity of the expression. The terms with Dirac deltas do not contribute to the optimization because for a finite cycle time, the magnetization doesn't ever saturate to the local steady state value, and the heat current is never completely zero.

Using these two equations, we can optimize the protocol $h(t)$ as before to obtain the maximum efficiency protocol corresponding to the PWC protocol touching the boundaries, as shown in the (top left) of Fig. \ref{var_eff_proto}. This result is consistent with what we demonstrated in subsection \ref{subsection:optimizing_efficiency}, thereby validating our method.

In Fig. \ref{var_eff_proto} (top left), we show the final form of the optimal efficiency protocol, which is piecewise constant, and the efficiency closely approaches the Carnot limit. Consequently, we plot the entropy production $\Delta S$ against the number of iteration steps. Based on the previous analysis, we anticipate nearly zero entropy production after optimization, indicating that the system is nearly reversible, as confirmed in Fig. \ref{var_eff_proto} (top right).
It is important to note that setting the ratios exactly equal, such that $\frac{T_h}{T_c} = \frac{h_{max}}{h_{min}}$, leads to numerical instability during the efficiency optimization, resulting in unphysical values of entropy change. Therefore, we use $h_{max} = 0.799$ and $h_{min} = 0.201$ in this section. The magnetization during the cycle is shown in Fig. \ref{var_eff_proto} (bottom left), while the magnetization versus field plot for the optimal efficiency protocol is presented in Fig. \ref{var_eff_proto} (bottom right). Note that the change in magnetization is very small, around $0.002$, as expected, since $\beta_h$ remains nearly constant. In the next Sec.  \ref{sec:Comparison_all_protocol}, we compare power and efficiency obtained using different methods discussed above.
\section{Comparisons of efficiency and power employing diverse methodologies.}
\label{sec:Comparison_all_protocol}
In the preceding sections (see Sec. \ref{section:Normal_protocol_long cycle time limit} and Sec. \ref{sec:optimization_and_optimal_protocol} and Sec. \ref{variational}), we explored various methods for the calculation of efficiency and for optimizing the efficiency and power using gradient descent algorithm and using the variational iterative method. Additionally, Table \ref{Table_II} presents a comparative analysis of efficiency values computed for four distinct protocols utilizing the previously mentioned approach and summarized below. 
\begin{table}[!t]
\begin{center}
\begin{tabular}{ 
|p{3.8cm}|p{1.2cm}|p{1.2cm}|p{1.2cm}|p{1.2cm}|p{1.4cm}|p{1.2cm}|p{1.2cm}|p{1.2cm}|}
 \hline
 \multicolumn{9}{|c|}{Efficiencies and Powers obtained through various mechanisms discussed in the text} \\
 \hline
 Protocol Type & $\eta$ & $\eta^*$ & $\eta_{num}$  & $\eta_{P^*}$ & $P^{*}$ ($\times10^{-3}$) & $\eta_{CA}$ &$\eta_{C}$ &$\eta_{C}^{*}$\\
 \hline
Piece-wise Constant  & $ 0.750$ & $0.750$  & 0.747  &0.436&  9.755& 0.500&0.750 &0.875\\
\hline
  Piece-wise Linear & $0.636 $ & $0.672$  & 0.748  &0.473& 14.967 & 0.500  & 0.750  &0.875 \\
 \hline
  Inverse Quadratic & $ 0.628$ & $0.665$  & 0.748  &0.472& 14.599 & 0.500 & 0.750  &0.875  \\
 \hline
  Sinusoidal  & $ 0.605$ & $0.643$   & 0.747  &0.436& 9.755& 0.500  & 0.750  &0.875  \\
 \hline
 Variational (hard) &\centering{---}&\centering{---}&0.749&0.475&16.322&0.500&0.750&---\\
 \hline
  Variational (soft) &\centering{---}&\centering{---}&0.714&0.472&16.042&0.500&0.750&--- \\
 \hline
\end{tabular}
%W: 0.00188482 η: 0.749223 for variational eff optim. Hard.
\caption{Table of efficiencies calculated using different methodologies. $\eta$ is the engine's efficiency and $\eta_{C}$ is the Carnot efficiency when $\dfrac{h_{min}}{h_{max}}=\dfrac{T_c}{T_h}$~ given,   ~($T_h=2.0, T_c=0.5$, $h_{max}=0.8$ and $h_{min}=0.2$) and $\eta^{*}$ is the efficiency and $\eta_{C}^{*}$ is the Carnot efficiency when $\dfrac{h_{min}}{h_{max}}\neq \dfrac{T_c}{T_h}$ given,~($T_h=4.0, T_c=0.5$, $h_{max}=0.8$ and $h_{min}=0.2$)~. ~$\eta_{num}$~ is the efficiency obtained by numerical algorithms: Gradient descent for the first four and numerical variational optimization for the last two, targeting maximum efficiency. We define the efficiency at maximum power by $\eta_P^*$. $\eta_{CA}$ is Curzon-Ahlborn efficiency. The parameters for all the protocols used for the first two columns are $a=0.8$, $b=0.4$, $c=0.2$, $d=0.3$, $r=0.5$, $\tau=12$, and $\mu=1.0$. The last two rows are for variational principle-based optimization, already discussed in sec. \ref{variational}. The value of the coefficient of the soft bounds is $0.5$ in power optimization and $5.0$ in efficiency optimization.}
\label{Table_II}
\end{center}
\end{table}
We calculate the values in the table using specific parameters for two different situations. We denote $ \eta$ to be the engine efficiency when ~$\dfrac{h_{min}}{h_{max}}=\dfrac{T_c}{T_h}$, with, $T_h=2.0, T_c=0.5$. Efficiency $\eta^{*}$~ and the Carnot efficiency ~$\eta_{C}^{*}$~ when ~$\dfrac{h_{min}}{h_{max}}\neq \dfrac{T_c}{T_h}$~ where, $T_h=4.0, T_c=0.5$. The protocol's initial parameters are $a=0.8$, $b=0.4$, $c=0.2$, $d=0.3$, ~$r=0.5$, ~$\tau=12$ and ~$\mu=1.0$. We determine the efficiency $\eta$ using  Eq. \eqref{eq:effi_all}. Unlike the earlier protocol discussed in Sec. \ref{sec:Analytics_Method} (Fig. \ref{protocolfig}), these protocols given in the table Table \ref{TableI}, contribute to the adiabatic work due to the jumps at $\tau/2$ and $\tau$. Thus, in these cases non-zero work contributions namely $W_3=W_4\neq 0$ occur. For the PWC protocol, efficiency reaches near the Carnot efficiency $\eta_C$ as we choose $\dfrac{h_{min}}{h_{max}}=\dfrac{T_c}{T_h}$. In contrast, for the other protocols mentioned in the second column, the efficiency falls below the Carnot efficiency since $\frac{h_{max}}{h_{min}} \neq \frac{T_c}{T_h}$. We calculate $\eta_{GD}$ using the Gradient descent optimization process by optimizing the efficiency as detailed in subsection \ref{subsection:optimizing_efficiency}. During efficiency optimization, all protocols converge towards the maximum efficiency protocol, with their values approaching the efficiency of PWC and hence the Carnot efficiency. Moreover, the efficiency calculated using the hard constraint in the variational principle method is higher than the results obtained from the other methodologies indicated in the last two rows of Table \ref{Table_II}. Another important quantity we discuss is the efficiency at maximum power denoted by $\eta_P^*$. We optimize the power, as discussed in subsection \ref{subsection:optimizing_power}. The optimal values of power are presented in the Table \ref{Table_II}. Notably, the Piecewise Linear and Inverse Quadratic protocols yield the highest power outputs, with their efficiencies close to the Curzon-Ahlborn efficiency, and matched with the values obtained from the formula $\eta_{CA}=1-\sqrt{\dfrac{T_c}{T_h}}$. During power optimization, the PWC and sinusoidal protocol converge to the same protocol which is also constant in each branch, but with different values than $h_{max}$ and $h_{min}$. This produces the maximum power achievable with their functions, but the efficiency values are lower than the PWL and inverse quadratic. This happens because, due to their shape restrictions, they are not able to extract maximum work possible, and a higher amount of input heat is wasted as heat. Notably, the optimal power value calculated using the variational principle is higher than the other methods. Overall, this approach of using the variational iterative method demonstrates superior optimization of efficiency and power for the spin system compared to the gradient descent algorithm.

\section{Discussion}
\label{section:Conclusion_futurework}
This paper presents a detailed study of a single spin experiencing a cyclically time varying external magnetic field. The spin is attached and interacts with two distinct heat reservoirs at different temperatures during different strokes of the cycle. Our analysis led to the derivation of an analytical expression for the magnetization, which, in turn, allows us to calculate various thermodynamic quantities for a general protocol. Notably, we observe that in the long cycle time limit, the average values rely solely on the initial and final point of the protocol. Furthermore, altering the ratio of the maximum and minimum value of the magnetic field can enhance the efficiency. We observe that as this ratio increases, efficiency rises but eventually saturates well below the corresponding Carnot efficiency. Additionally, we demonstrate the application of well-established optimization techniques to enhance the system's efficiency. Subsequently, we optimize the power output and compute the efficiency at maximum power, known as the Curzon-Ahlborn efficiency, using a gradient descent algorithm for four distinct protocols. Much research has consistently highlighted the Piece-wise Constant (PWC) protocol as the optimum efficiency protocol. We show that all protocols converge towards the PWC protocol during the efficiency optimization. However, when optimizing for maximum power output, the optimal protocol is not the piece-wise constant. Therefore, the PWL and the inverse-quadratic protocols were able to perform much better. Our study conducts a comparative analysis of efficiency and power results across different protocols and timescales. We then employ an iterative variational principle-based algorithm to derive the optimal power protocol. This method is validated by deriving a piecewise constant protocol for maximum efficiency and analyzing the system's response to this optimal power protocol. We also examine and compare the entropy production during the optimization of both power and efficiency, as illustrated in (Fig. \ref{mag_vs_h}~(bottom right)). The figure demonstrates that the maximum power protocol results in substantial entropy production, leading to significant energy dissipation as heat and a reduction in efficiency. This highlights the trade-off between efficiency and power: achieving maximum efficiency necessitates large field changes to counteract magnetization changes caused by the temperature fluctuations. In the first half of the cycle, a significant decrease in the field is not feasible, if the field must continuously decrease, while a constant field results in no work being performed. This explains the differences in the shapes of the optimal protocols for the two distinct objectives. Overall this approach of using variational iterative method showcases superior optimization of efficiency and power for a spin system compared to the gradient descent algorithm. We also note that the forms of optimal protocols, for both power and efficiency, obtained using these techniques are simple and can be realized experimentally. Application of the variational method to cycle time optimization as well as the temperature protocol optimization and applications to other microscopic engines in currently underway.

\section{Acknowledgment}
Authors gratefully acknowledge Science and Engineering Research Board (SERB), India for financial support through the MATRICS grant (No. MTR/2020/000349). Ri. M. and Ra. M. acknowledge Sandwich Training Educational Program (STEP) by The Abdus Salam International Center for Theoretical Physics (ICTP), Trieste, Italy where a part of this work was done. M. C. gratefully acknowledges the University Grants Commission (UGC), India for the financial support through the junior research fellowship number $23161018714$. Authors thank Vipul Upadhyay and Bijay Kumar Agarwalla for useful discussions. Authors also thank Viktor Holubec, for several discussions and especially for pointing out references \cite{CraunOptimalJDyn15} and \cite{Symmetry} which were very crucial for this study.

\newpage
\appendix
\counterwithin{figure}{section}
\section{Pseudo-codes for the optimization algorithms}
\subsection{Using gradient descent} \label{appendix_GD}
In Sec. \ref{Optimization_using_GD}, we discuss the method to optimize the control of the field to make it work as an engine. To achieve this, we fit four-parameter functions to achieve maximum power or efficiency. The following is an example of pseudo-code to describe the algorithm used to optimize work for a fixed cycle time: 
\begin{tcolorbox}[title=\centering Pseudo-code for gradient descent based optimization of work, fontupper=\small]\label{GDalgo}
\begin{lstlisting}[escapechar=']
//control boundaries and temperatures of baths.
#define Th, Tc, hmax, hmin; 
initialize(params);
W = work(params);//should be redefined for each protocol type (e.g. for PWC)
converged=false;
grad=zeros(size(params));
#define alpha,eps //learning rate and tolerance for convergence.
while(not converged):
    new_params=params;
    for i = 1 to size(params):
        new_params[i] = params[i] + h;
        W_p=work(new_params);
        new_params[i] = params[i]-h;
        W_m=work(new_params);
        grad[i]=(W_p-W_m)/2h;
    end for
    params+=alpha*grad; \\ Using Eq. '\ref{updation_equation}'
    Wnew=work(params);
    if (norm(grad) < eps) then converged=true;
    W=Wnew;
end while

print("Optimum params are: ",params);
\end{lstlisting}
\end{tcolorbox}
We vary all the protocol parameters individually, and we define the gradient vector based on the partial derivatives of work w.r.t each parameter. Then, we update them together until the magnitude of the gradient vector becomes smaller than a threshold.
\subsection{ Using variational principle} \label{Varalgo}
Next, we use a different method based on the variational principle to produce optimal protocols through an iterative process. The pseudo-code to describe this is as follows,
\begin{tcolorbox}[title=\normalsize{Pseudo-code to obtain maximum power protocol using the Variational Principle},fontupper=\small]
\begin{lstlisting}[escapechar=']
#define tau, dt, Th, Tc;   Let N=tau/dt; 
#define m,Temp,h,Lambda=zeros(N);
initialize(m,h); //initialize these arrays to some smooth functions.
Temp = join(Th*ones(N/2), Tc*ones(N/2));
#define eps; //tolerance for protocol convergence
converged=false;
W=work(m,h,Temp);//value of work from the current system state
#declare Lambda_dot,m_dot,del_h; //functions of m,h,Temp and Lambda.
while(not converged):
    m[0]=m[N-1];
    for i=0 to N-1: //repeat to ensure periodicity
    // Discretization of Eq. '\ref{eq_dyna}'
        m[i+1] = m[i] + m_dot(m[i],h[i],Temp[i])*dt; 
    end for
    Lambda[N-1]=Lambda[0];//repeat to ensure periodicity
    for i=N-1 to 0:
    //Discretization of Eqs. '\ref{eq_lambda_pow}' and '\ref{eq_lambda_eff}' 
        Lambda[i-1] = Lambda[i] - Lambda_dot(Lambda[i],h[i])*dt;
    end for
    for i=0 to N-1:
    \\Discretization of Eqs. '\ref{eq_h_pow}' and '\ref{eq_h_eff}'
        h[i+1] = h[i] + del_h(m[i],h[i],Temp[i],Lambda[i]);
        if(h[i+1]>hmax) then h[i+1]=hmax;
        if(h[i+1]<hmin) then h[i+1]=hmin;
    end for
    Wnew=work(m,h,Temp);
    if (Wnew-W<eps) then converged=true;
    W=Wnew
end while
plot(h);
\end{lstlisting}
\end{tcolorbox} 
We iterate to determine the optimal protocol after initializing the field with a smooth function. We calculate the magnetization in each iteration, derive the corresponding $\lambda(t)$ function for $m(t)$ and $h(t)$, and then update the field by $\delta h(t)$. The increment of the field $h(t)$ depends on the type of boundary conditions applied. An additional step is necessary for hard boundary conditions (lines $23$ and $24$). On the other hand, soft boundary conditions incorporate an extra restoring force term directly into the $\delta h(t)$ calculation.
\section{Exploring the effect of cycle time and softness constraints on protocol shape}\label{cycle_time}
\begin{figure}[!t]
    \centering    
    \includegraphics[width=16cm, height=16cm, angle=0]{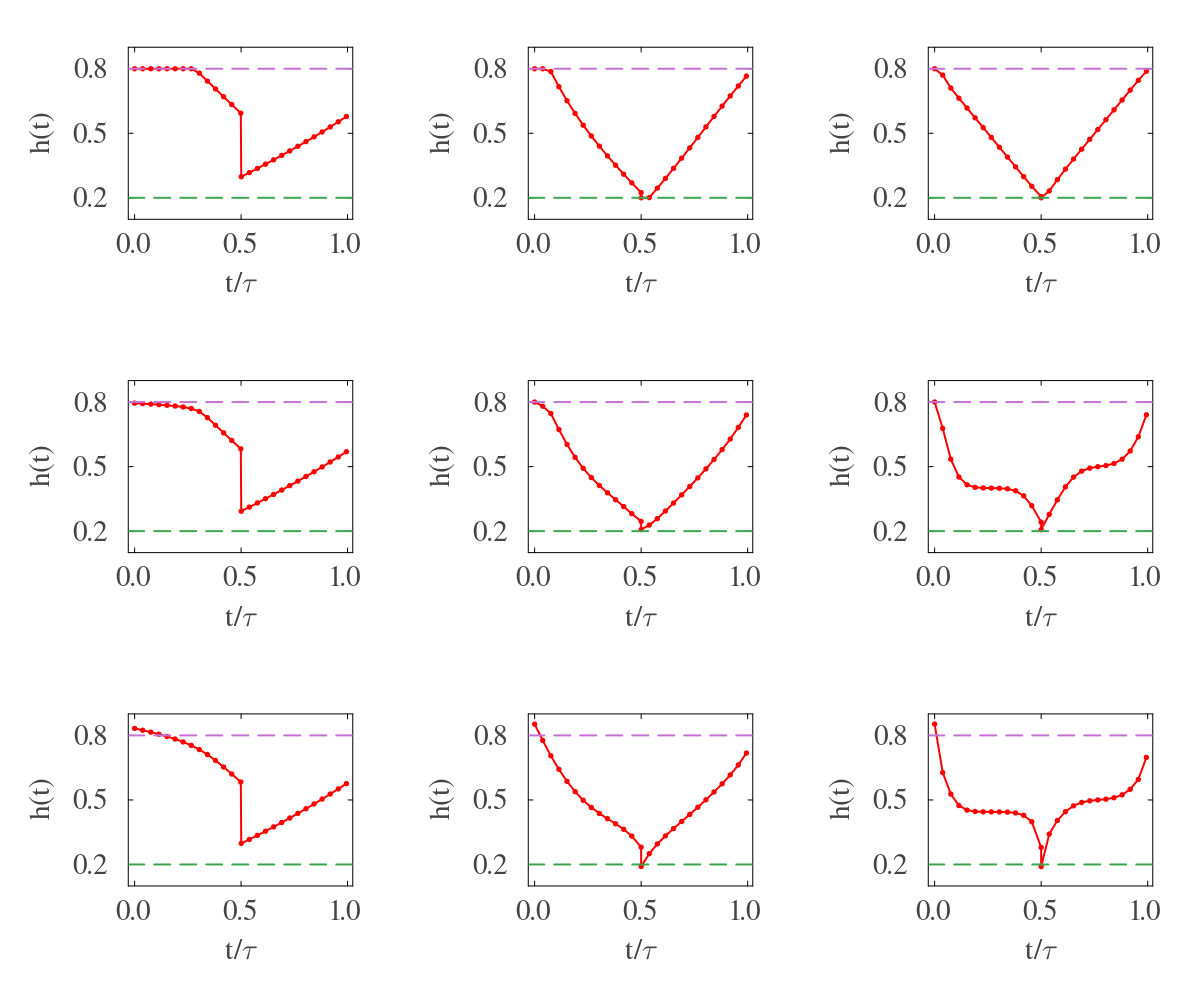}
    \caption{Represents optimal power protocols. The columns correspond to cycle times $\tau=5,\tau=50$, and $\tau=200$ respectively.   The first row is for hard constraints. The second row is for soft constraints with parameters $C_0=0.005, C_1=100$). The third row corresponds to soft constraints with parameters ($C_0=0.005,C_1=25$).} 
    \label{nineplots}    
\end{figure}

As discussed at the end of subsection \ref{softhard}, in power optimization, the shape of the protocol is very much dependent on the total cycle time. The jumps depend on the time given for relaxation, so the jumps are smaller in the high $\tau$ limit, as shown in Fig. \ref{nineplots}. One possible explanation for the decrease in jumps due to increased cycle time is as follows: During the jump at $\tau/2$, work is done on the system. At this point, the magnetization is at its lowest. At the jump of $\tau$, the magnetization value is at its highest, and work is extracted at this point. So, the net work is extracted from the system. As the cycle time increases, the magnetization values can approach their equilibrium more closely, resulting in higher maxima and lower minima. Consequently, the same effect can be achieved with smaller jumps. Additionally, there are two parameters, $C_0$ and $C_1$, associated with the soft constraints. $C_0$ regulates the overall strength of the restoring force, while $C_1$ determines the 'hardness' of the boundary. A higher value of $C_1$ makes the boundary more similar to a hard boundary condition, whereas a lower value permits the protocol to slightly exceed it, allowing for a smoother shape.


\begin{thebibliography}{99}


\bibitem{sekimoto98} K. Sekimoto, \href{https://doi.org/10.1143/PTPS.130.17}{ Prog. of Theoretical Phys. \textbf{130}, 17–27 (1998).} 


\bibitem{Seifert12} U. Seifert, \href{https://dx.doi.org/10.1088/0034-4885/75/12/126001}{Rep. Prog. Phys. \textbf{75}, 126001 (2012).}

\bibitem{Martinez17} I. A. Martinez, E. Roldan et al., \href{https://doi.org/10.1039/C6SM00923A}{Soft Matter \textbf{13}, 22 (2017).}

\bibitem{Marathe18} A. Saha, R. Marathe, et al., \href{https://dx.doi.org/10.1088/1742-5468/aae84a}{ J. Stat. Mech. Theory Exp. \textbf{2018}, 113203 (2018).}

\bibitem{Marathe19} A. Saha and R. Marathe, \href{https://dx.doi.org/10.1088/1742-5468/ab39d4}{ J. Stat. Mech. Theory Exp. \textbf{2019}, 094012 (2019).}

\bibitem{Marathe22} R. Majumdar, A. Saha and R. Marathe, \href{https://dx.doi.org/10.1088/1742-5468/ac7e3d}{J. Stat. Mech. Theory Exp. \textbf{2022}, 073206 (2022).}

\bibitem{Rana14} S. Rana, P. S. Pal, A. Saha et al., \href{https://link.aps.org/doi/10.1103/PhysRevE.90.042146}{Phys. Rev. E \textbf{90}, 042146 (2014).}

\bibitem{Ruben23} C. Antonio, G.Valadez et al., \href{https://doi.org/10.1016/j.physa.2022.128342}{Physica A \textbf{609}, 128342 (2023).}

\bibitem{Glauber} R. Glauber, \href{https://doi.org/10.1063/1.1703954}{J. Math. Phys. \textbf{4}, 294-307 (1963).}

 \bibitem{Marathe05} R. Marathe, A. Dhar, \href{https://doi.org/10.1103/PhysRevE.72.066112}{Phys. Rev. E \textbf{72}, 066112 (2005).}

 \bibitem{Marathe07} R. Marathe, A. M. Jayannavar et al., \href{https://doi.org/10.1103/PhysRevE.75.030103}{Phys. Rev. E \textbf{75}, 030103 (2007).}

\bibitem{Marathe17} D. Basu,  J. Nandi et al., \href{https://doi.org/10.1103/PhysRevE.95.052123}{Phys. Rev. E  \textbf{95}, 052123 (2017).}

\bibitem{Chen02} J. Chen, B. Lin et al., \href{https://dx.doi.org/10.1088/0022-3727/35/16/322}{J. Phys. D: Appl. Phys. \textbf{35}, 2051 (2002).}

\bibitem{Myers22} N. M. Myers, O. Abah, et al., \href{https://doi.org/10.1116/5.0083192}{AVS Quantum Sci. \textbf{4}, 027101 (2022).}

\bibitem{Bechinger12} V. Blickle and C. Bechinger, \href{https://doi.org/10.1038/nphys2163}{Nat. Phys. \textbf{8}, 143 (2012).}

\bibitem{Martinez16} I. A. Martinez, E. Roldan et al., \href{https://doi.org/10.1038/nphys3518}{Nat. Phys, \textbf{12}, 67 (2016).}

\bibitem{Krishnamurty16} S. Krishnamurthy, S. Ghosh et al., \href{https://doi.org/10.1038/nphys3870}{Nat. Phys. \textbf{12}, 1134 (2016).}

\bibitem{Albay21} J. A. C Albay,  Z.-Y- Zhou et al., \href{https://doi.org/10.1038/s41598-021-83824-7}{Sci. Rep. \textbf{11}, 4394 (2021).}

\bibitem{Cheng22} K. Cheng, P. Liu et al., \href{https://doi.org/10.1039/D1SM01798E}{Soft Matter \textbf{18}, 2541 (2022).}

\bibitem{Robnagel16} J. Ro{\ss}nagel,  S. T. Dawkins et al. \href{https://doi.org/10.1126/science.aad6320}{Science \textbf{352}, 325 (2016).}

\bibitem{Assis19} R. J. de Assis, T. M. de Mendonça et al,  \href{https://link.aps.org/doi/10.1103/PhysRevLett.122.240602}{Phys. Rev. Lett. \textbf{122},  240602 (2019).}

\bibitem{Klimovsky18} D. G. Klimovsky,  A. Bylinskii et al.,  \href{https://doi.org/10.1103/physrevlett.120.170601}{Phys. Rev. Lett. \textbf{120}, 170601 (2018).}

\bibitem{Barontini19} G. Barontini and M. Paternostro, \href{https://dx.doi.org/10.1088/1367-2630/ab2684}{New J. Phys. \textbf{21}, 063019 (2019).}

\bibitem{Levy20} A. Levy, M. Göb et al., \href{https://doi.org/10.1088/1367-2630/abad7f}{New J. Phys.  \textbf{22}, 093020 (2020).}

\bibitem{Peterson19} J. P. S. Peterson, T. B. Batalhão, et.al, \href{https://doi.org/10.1103/PhysRevLett.123.240601}{Phys.Rev.E  \textbf{123}, 240601 (2019).}

\bibitem{Abah12} O. Abah, J. Roßnagel et al., \href{https://doi.org/10.1103/PhysRevLett.109.203006}{Phys. Rev.Lett. \textbf{109}, 203006 (2012).}

\bibitem{Bera22} M. L. Bera, S. J. Farré, \href{https://doi.org/10.1103/PhysRevResearch.4.013157}{Phys. Rev. Research \textbf{4}, 013157, (2022).}

\bibitem{Xiao14} G. Xiao and J. Gong \href{https://doi.org/10.1103/physreve.90.052132}{Phys. Rev. E \textbf{90}, 052132 (2014).}

\bibitem{Abiuso20} P. Abiuso and M. P. Llobet \href{https://doi.org/10.1103/PhysRevLett.124.110606}{Phys. Rev. Lett. \textbf{124} 110606 (2020).}

\bibitem{Seifert07} T. Schmiedl and U. Seifert, \href{https://doi.org/10.1103/PhysRevLett.98.108301}{Phys.Rev.Lett. \textbf{98}, 108301 (2007).}

\bibitem{Udo08} T. Schmiedl and U. Seifert, \href{https://doi.org/10.1209/0295-5075/81/20003}{EPL \textbf{81}, 20003 (2008).}

\bibitem{Puglisi21} G. Gronchi and A. Puglisi, \href{https://doi.org/10.1103/PhysRevE.103.052134}{Phys. Rev. E \textbf{103}, 052134 (2021).}

\bibitem{Holubec18} V. Holubec, and A. Ryabov, \href{https://doi.org/10.1103/PhysRevLett.121.120601}{Phys. Rev. Lett. \textbf{121}, 120601 (2018).}

\bibitem{Ye22} Z. Ye and F. Cerisola et al., \href{https://doi.org/10.1103/PhysRevResearch.4.043130}{Phys. Rev. Res. \textbf{4}, 043130 (2022).}

\bibitem{Fu2023} J. F. Chen, H. T. Quan, \href{https://doi.org/10.48550/arXiv.2310.19622}{arXiv:2310.19622 (2023).}

\bibitem{Dechant16} A. Dechant, N. Kiesel et al., \href{https://dx.doi.org/10.1209/0295-5075/119/50003}{EPL. \textbf{119}, 50003 (2017).}

\bibitem{MassimoPRL10} M. Esposito, R. Kawai et al., \href{https://doi.org/10.1103/PhysRevLett.105.150603}{ Phys. Rev. Lett. \textbf{105}, 150603 (2010).}

\bibitem{Massimo10} M. Esposito, R. Kawai, et al., \href{https://doi.org/10.1103/PhysRevE.81.041106}{Phys. Rev. E \textbf{81}, 041106 (2010).}

\bibitem{Solon18} A. P. Solon and J. M. Horowitz, \href{https://doi.org/10.1103/PhysRevLett.120.180605}{Phys. Rev. Lett. \textbf{120}, 180605 (2018).}

\bibitem{Erdman23} P. A. Erdman, A. Rolandi et al. \href{https://doi.org/10.1103/physrevresearch.5.l022017}{Phys. Rev. Research \textbf{5}, L022017 (2023).}

\bibitem{Plata19}C. A. Plata, D. Guéry-Odelin et al., \href{https://doi.org/10.1103/PhysRevE.99.012140}{Phys. Rev. E \textbf{99}, 012140 (2019).}

\bibitem{Curzon} F. L. Curzon,  B. Ahlborn, \href{https://doi.org/10.1119/1.10023}{Am. J. Phys. \textbf{43}, 22 (1975).}

\bibitem{Ignatio16} I. A. Martinez, A. Petrosyan, et al., \href{https://www.nature.com/articles/nphys3758}{Nat. Phys. \textbf{12}, 843 (2016).}

\bibitem{Zhang} Y.  Zhang, \href{https://doi.org/10.1007/s10955-020-02508-0}{J. Stat. Phys. \textbf{178}, 1336 (2020).}

\bibitem{Abiuso22} P. Abiuso, V. Holubec, et al., \href{https://dx.doi.org/10.1088/2399-6528/ac72f8}{J. Phys. Commun. 6, 063001 (2022).}

\bibitem{Zhong22} A. Zhong and M. R. DeWeese, \href{https://doi.org/10.1103/PhysRevE.106.044135}{Phys. Rev. E \textbf{106}, 044135 (2022).}

\bibitem{Seifert16} M. Bauer, K. Brandner, and U. Seifert, \href{https://doi.org/10.1103/PhysRevE.93.042112}{Phys. Rev. E \textbf{93}, 042112 (2016)}

\bibitem{Chen22} Y. H. Chen et al., \href{https://doi.org/10.1103/PhysRevE.106.024105}{Phys. Rev. E \textbf{106}, 024105 (2022).}

\bibitem{CraunOptimalJDyn15} M. Craun and B. Bamieh, \href{https://doi.org/10.1115/1.4029682}{J. Dyn. Sys., Meas., Control \textbf{137}, 071002 (2015).}

\bibitem{Symmetry} R. Paul, K. H. Hoffmann, \href{https://doi.org/10.3390/sym13050873}{Symmetry, \textbf{13}, 873 (2021).}

\end{thebibliography}
\end{document}